\documentclass[11pt,a4paper]{article}
\usepackage{jinstpub}

\usepackage[utf8]{inputenc} 
\usepackage[T1]{fontenc}    
\usepackage{url}            
\usepackage{booktabs}       
\usepackage{amsfonts}       
\usepackage{nicefrac}       
\usepackage{microtype}      
\usepackage{lipsum}
\usepackage{xspace}
\RequirePackage{siunitx}
\RequirePackage{bm}

\def\Xmax{$X_\mathrm{max}$\xspace}

\title{Expected performance of air-shower measurements with the radio-interferometric technique}

\author[a,b,1]{F.~Schlüter\note{Corresponding author.}}
\author[a,c]{T.~Huege}
\affiliation[a]{Karlsruher Institut für Technologie, Institut für Astroteilchenphysik, Karlsruhe, Germany}
\affiliation[b]{Universidad Nacional de San Martín, Instituto de Tecnologías en Detección y Astropartículas,\\ Buenos Aires, Argentina}
\affiliation[c]{Vrije Universiteit Brussel, Astrophysical Institute, Brussels, Belgium}
\emailAdd{felix.schlueter@kit.edu}

\abstract{
Interferometric measurements with arrays of radio antennas are a powerful and widely used technique in astronomy. Recently, this technique has been revisited for the reconstruction of extensive air showers \cite{schoorlemmer2020radio}. This ``radio-interferometric technique'' exploits the coherence in the radio emission emitted by billions of secondary shower particles to reconstruct the shower parameters, in particular the shower axis and depth of the shower maximum \Xmax. The accuracy previously demonstrated on simulations with an idealized detector is very promising. The prospect of an accurate \Xmax measurement for inclined air showers combined with measurements of the electromagnetic energy (also with radio antennas) and the muonic shower content (via ground particle detectors) is very intriguing as it would provide a high sensitivity to the mass of cosmic rays, key information to study their origin. In this article we evaluate the potential of interferometric \Xmax measurements using (simulated) inclined air showers with sparse antenna arrays under realistic conditions. To determine prerequisites for the application of the radio-interferometric technique with various antenna arrays, the influence of inaccuracies in the time synchronisation between antennas and its inter-dependency with the antenna density is investigated in detail. We find a strong correlation between the antenna multiplicity (per event) and the maximum acceptable time jitter, i.e., inaccuracy in the time synchronisation. For data recorded with a time synchronisation accurate to within \SI{1}{ns} in the commonly used frequency band of \SIrange{30}{80}{MHz}, an antenna multiplicity of $\,> 50$ is needed to achieve an \Xmax resolution of $\sigma_{X_\mathrm{max}} \lesssim 20\,$g$\,$cm$^{-2}$. For data recorded with \SI{2}{ns} accuracy, already $\gtrsim 200$ antennas are needed to achieve this \Xmax resolution. Furthermore, we find no advantage reconstructing \Xmax from data simulated at higher observation frequencies, i.e., up to several hundred MHz. Finally, we provide a generalisation of our results from very inclined air showers to vertical geometries. 
}


\begin{document}
\maketitle

\section{Introduction}
Radio signals from extensive air showers exhibit wave phenomena. An example for such phenomena is the nowadays well-established Cherenkov cone which results from the interference of the signals emitted by billions of shower particles in an atmosphere with a refractive index gradient.

Interferometric techniques expose this coherence in the radio emission. Thereby, both the signal's amplitude and phase information is used, while traditional reconstruction methods of extensive air showers rely on the amplitude information only. Interferometric techniques are standard in radio astronomy, where sources are at infinity and hence all antennas receive the same signal with a planar wavefront. Application to radio emission from extensive air showers is more challenging, as the source is typically nearby, is extended, and the emission from different parts of the shower propagates through different refractive index gradients.

Interferometric techniques have previously been used successfully for cosmic-ray radio detection in the LOPES experiment to identify coherent air-shower radio pulses amongst strong and time-correlated radio-frequency interference, to estimate the energy of the primary particle, and to provide an image of the intensity distribution on the sky from which the arrival direction can be determined  \cite{Falcke:2005tc}. They have also been employed to determine the depth of shower maximum from LOPES data \cite{Apel:2014usa,Apel:2021oco} with an experimental accuracy worse than 100 g/cm$^2$ but potential in pure simulations to reach an accuracy as good as 30 g/cm$^2$. Attempts to apply interferometric techniques to ground-based radio arrays with a larger extension than the small-scale LOPES experiment, for example within the Auger Engineering Radio Array, had not been successful \cite{Jandt}, presumably because the then-made assumption that antennas see identical signals no longer holds for larger arrays. Another experiment routinely using interferometric techniques to identify and reconstruct air-shower radio emission is ANITA \cite{Romero-Wolf:2014pua}. Finally, real-time interferometric triggering is also being investigated for particle showers in ice \cite{Vieregg:2017ntx} and air \cite{Hughes:2020ghq}.

In \cite{schoorlemmer2020radio}, the so-called radio-interferometric technique (RIT) is developed and successfully applied to air shower simulations with an idealized detector (zenith-angle dependent dense antenna array, perfect time synchronisation between antennas and perfectly known antenna locations) to reconstruct the shower axis and depth of the shower maximum \Xmax with high accuracy. A resolution of better than \SI{0.04}{\degree} (< \SI{0.2}{\degree}) in the arrival direction and \SI{3}{g\,cm^{-2}} (\SI{10}{g\,cm^{-2}}) in \Xmax for inclined (vertical) showers is demonstrated.

The limited size of the footprint illuminated by the strongly forward-beamed radio emission in vertical air showers demands the use of comparatively dense and small antenna arrays and thus restricts the observation of cosmic rays to energies around and below \SI{1}{EeV} \cite{Huege:2016veh, Schroder:2016hrv}. In inclined air showers the radio-emission footprint is spread over large areas thus enabling the observation of air showers with sparse antenna arrays \cite{HuegeUHECR2014, Aab:2018ytv}. This allows one to instrument large areas (> \SI{1000}{km^2}) to detect ultra-high energy cosmic rays (UHECRs) with energies up to $\sim \,$\SI{100}{EeV}, soon to be realized with the AugerPrime Radio Detector \cite{2019ICRC_pont}. An accurate reconstruction of \Xmax for inclined air showers using RIT in addition to the measurement of the energy content of the electromagnetic cascade by the same radio antennas and the mounic content by ground-particle detectors would provide excellent sensitivity to the mass composition of cosmic rays \cite{Holt:2019fnj, Frank:Lol} and could thus provide key information in the quest for the origin of UHECRs . 

Here, we investigate whether the promising results achieved in \cite{schoorlemmer2020radio} for simulations with an idealized detector (zenith-angle dependent dense antenna array, perfect time synchronisation between antennas) can be confirmed for air showers measured with realistically dimensioned air shower detector arrays, i.e., coarse discretely spaced antenna\footnote{With ``antenna'' we refer to an antenna(-station) consisting of at least two orthogonally aligned antennas allowing to determine the full 3-dimensional electric field of the incoming radio emission. In the context of air-shower simulations ``antenna'' refers to a location at which the radio pulse
is sampled.}
arrays, as needed to instrument the required large fiducial areas, and an imperfect time synchronisation between antennas not connected by cables. The primary objective of this study is to investigate the application of RIT for inclined air showers, where the potential is largest in terms of achievable \Xmax resolution and complementarity to ground-based measurements, and to formulate prerequisites for the application of RIT with sparse antenna arrays which can cover the required large fiducial areas. 

For interferometry, the signal arrival times and the antenna positions have to be known very accurately to preserve the coherence within the measured signals. In \cite{schoorlemmer2020radio}, the authors quote that the timing accuracy has to be better than a quarter of the signals' oscillation period, e.g., $\sigma_t = \sqrt{\sigma_{t_\mathrm{signal}} ^ 2 + (\sigma_{\vec{x}_\mathrm{antenna}} / c) ^ 2} < (4 \cdot \nu)^{-1} \sim \,$\SI{5}{ns} at a frequency of $\nu = 50\,$MHz. Furthermore they report that a maximum inaccuracy of $\sigma_t = 3\,$ns yields accurate results. In \cite{SCHRODER2010277} a much more restrictive coherence criterion for the same frequency band is concluded: a twelfth of the period or $\sigma_t <\,$\SI{1}{ns} at \SI{80}{MHz} (this corresponds to \SI{1.667}{ns} at \SI{50}{MHz}). Air shower experiments which aim to instrument large areas rely on self-powering detector stations with wireless communication. Thus the time synchronisation between those stations, achieved with GPS clocks, is typically of the order of a few nanoseconds ($\sigma_t \sim\,$ \SIrange{5}{10}{ns}) \cite{2005ICRC....8..307A}. However, with specialized hardware such as a phase-stable beacon transmitter this might improve to the order of a nanosecond ($\sigma_t \lesssim 1\,$ns) \cite{SCHRODER2010277, AERA_TIME_2016}. The antenna positions can be determined within $\sim \,$\SI{10}{cm} with differential GPS surveys. Thus for measurements of the radio emission below $\lesssim \,$\SI{100}{MHz} the contribution of $\sigma_{\vec{x}_\mathrm{antenna}}$ to $\sigma_t$ can be ignored. However, for frequencies of several hundred MHz the $\sigma_{\vec{x}_\mathrm{antenna}}$ can become significant. Thus verifying which coherence criterion is sufficient is crucial for the design and planning of an experiment which aims to employ interferometric reconstructions. 

The investigation presented here mainly refers to the frequency band of the radio emission from \SIrange{30}{80}{MHz}. This frequency band, also used in \cite{schoorlemmer2020radio}, is used by most current-generation large scale radio detector arrays \cite{Holt:2017dyo, Schellart:2013bba, Bezyazeekov:2015rpa} as well as the upcoming AugerPrime Radio Detector \cite{2019ICRC_pont}. Additionally, we investigate the performance achievable with higher frequency bands, in particular \SIrange{50}{200}{MHz} as proposed for the GRAND experiment \cite{Alvarez-Muniz:2018bhp} and \SIrange{150}{350}{MHz} for even higher frequencies such as those accessible by the upcoming SKA-Low \cite{Huege:2015jga,deLeraAcedo2015} array or the IceCube Radio Surface Array \cite{schroder2019science}. Furthermore, we investigate how an inaccurate knowledge of the atmospheric refractivity profile affects the reconstruction. We do not consider ambient noise, i.e., radio-frequency-interference, in our study. However, we briefly discuss this matter in Sec.\ \ref{sec:diss}.

This article is structured as follows. First, we elaborate on the shower simulations used in this work. In section \ref{sec:rit} we describe the reconstruction of the shower axis and \Xmax with RIT. Furthermore in Sec.\ \ref{sec:atm} the effect of inaccuracies in the knowledge of the atmospheric refractivity on the reconstruction is shown. In section \ref{sec:auger} we evaluate RIT for inclined air showers with different zenith angles measured with a \SI{1.5}{km}-spaced antenna array. The effect of an inaccurate time synchronisation between antennas for different detector layouts, i.e., antenna arrays with different spacings is investigated in Sec.\ \ref{sec:timejitter}. In Sec.\ \ref{sec:freqs} the reconstruction in the higher frequency bands is evaluated. Finally we discuss the obtained results in Sec.\ \ref{sec:diss} and conclude in Sec.\ \ref{sec:conc}.

\section{Simulations}
\label{sec:simulations}
\begin{figure}[tbp]
    \centering
    \includegraphics[width=.85\linewidth]{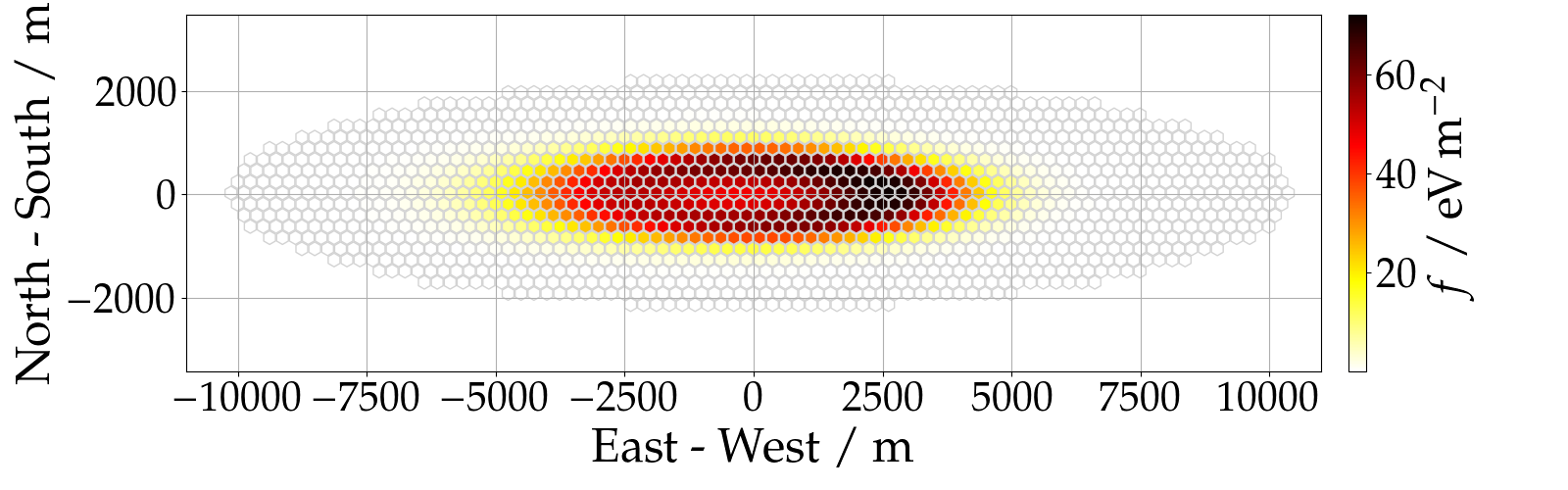}\\
    \includegraphics[width=.85\linewidth]{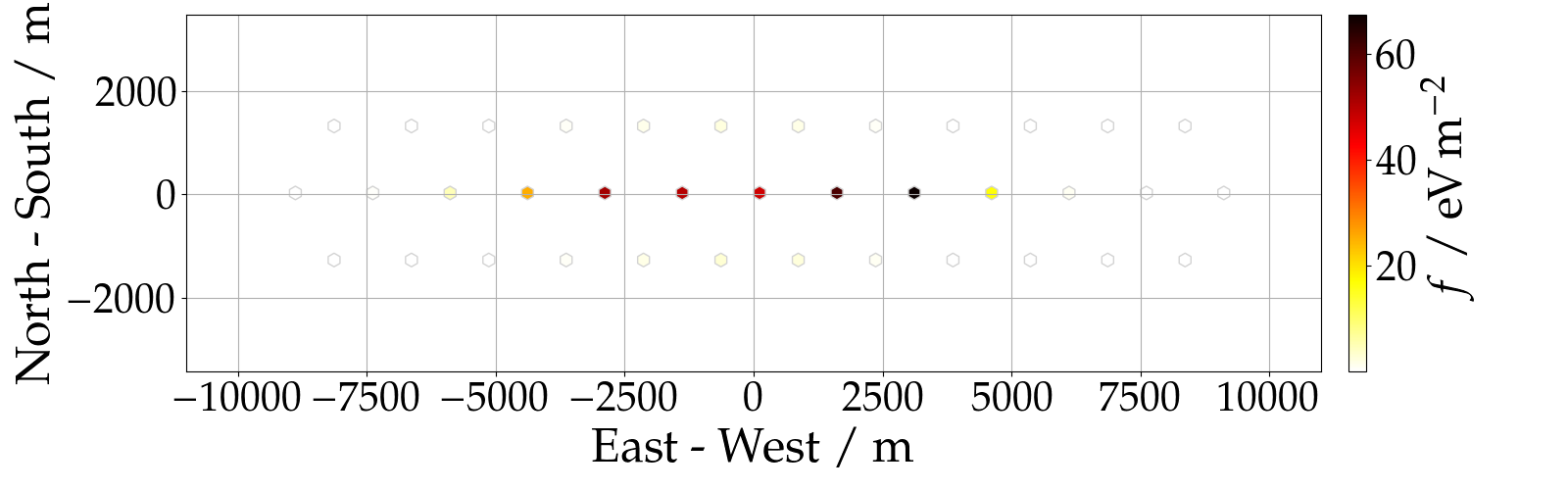}\\
    \includegraphics[width=.85\linewidth]{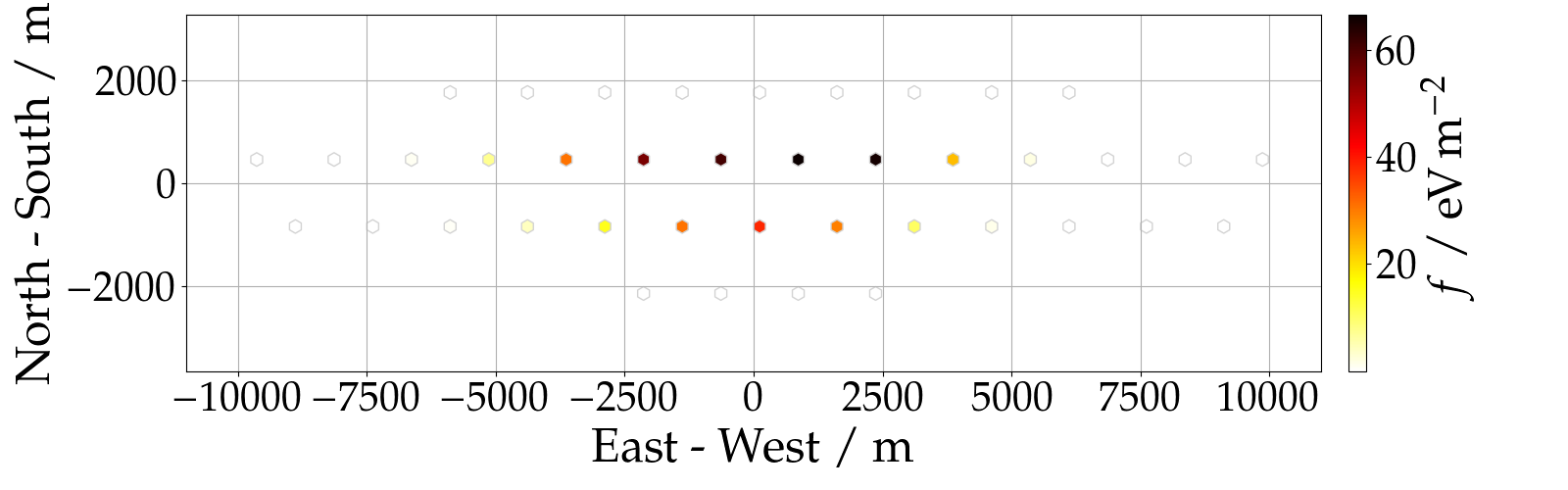}
    \caption{{\it Top}: Radio-emission footprint of a \SI{77.5}{\degree} zenith-angle air shower coming from east measured with a dense \SI{250}{m} array. The energy fluence $f$, i.e., energy deposit per square meter, is color coded. The footprint exhibits the typical Cherenkov cone. {\it Middle}: Same shower measured with a sparse \SI{1500}{m} (sub-)array with the same central antenna. {\it Bottom}: Same shower measured on a \SI{1500}{m} (sub-)array with a different central antenna.}
    \label{fig:array}
\end{figure}
We evaluate the potential of RIT using CoREAS \cite{Huege:2013vt} simulations. The simulations used in this work can be divided into two sets. The first set contains a total of 1902 showers, half of which is induced by proton and the other half by iron primaries. The showers are simulated with antennas situated on a hexagonal grid with \SI{1.5}{km} spacing, which corresponds to the configuration of the Pierre Auger Observatory and its upcoming large-scale radio detector. The second set contains 50 proton showers simulated on a very dense hexagonal grid with an antenna spacing of \SI{250}{m}. 

All simulations are performed with CORSIKA/CoREAS in version v7.7401 and, without loss of generality, for the ambient conditions of the Pierre Auger Observatory in October ($\equiv$ atmospheric profile, as listed in \cite[p. 162]{corsika}, with a refractive index at sea level of $n_0 = 1 + 3.12 \cdot 10^{-4}$), a magnetic field with an inclination of $\sim\,$\SI{-36}{\degree} and a strength of \SI{0.24}{\mu G}, and an altitude of the detector of \SI{1400}{m} a.s.l.. The chosen refractive index at sea level reflects the yearly average for the location of the Pierre Auger Observatory. The yearly fluctuations in refractivity are of the order of \SI{7}{\%} \cite[p. 51]{PhDGlaser}. We use QGSJetII-04 \cite{PhysRevD.83.014018} and UrQMD \cite{Bleicher_1999} as high- and low-energy hadronic interaction models and set a thinning level of \num{1e-6} with optimized weight limitation \cite{Kobal2001}.

The showers simulated with the \SI{1.5}{km} hexagonal grid cover the energy range between $\log(E / \text{eV}) = 18.4$ and $\log(E / \text{eV}) = 20.1$ uniformly randomized in $\log(E / \text{eV})$. The arrival directions, i.e., the azimuth $\phi$ and zenith $\theta$ angles, are uniformly randomized in $\phi$ from \SIrange{0}{360}{^\circ} and in $\sin^2\theta$ from \SIrange{65}{85}{^\circ}. The shower impact point at ground (in the following called ``core'') is randomly distributed within a finite \SI{3000}{km^2} detector array. For each shower all antennas are simulated within a maximum distance to the shower axis, beyond which the signals become negligible and are typically dominated by ambient radio-frequency background. The average number of simulated antennas per shower and the maximum antenna-axis distance binned in zenith angle are listed in Tab.\ \ref{tab:sim2}. For (actual) measurements the antenna multiplicity is, in addition to the detector layout and shower size (energy, arrival direction), also governed by the data acquisition system of the experiment, in particular the trigger determining which radio antennas to read out. In section \ref{sec:diss} the effect of (external) triggering is discussed.

To study the reconstruction performance for different detector layouts, i.e., array spacings, simulations with a very {\it dense} grid, which can be divided in several sub-arrays with larger antenna spacings, are suitable. Since the computational cost for each shower scales almost linearly with the number of simulated pulses we need to limit our phase space of densely sampled, simulated showers. Thus we simulate 50 proton showers with only one energy $\log(E / \text{eV}) = 18.4$, one zenith angle $\theta = 77.5^\circ$ and two azimuth angles $\phi = 0^\circ$ (arriving from geomagnetic east) and $\phi = 30^\circ$ (arriving from north of east), for each of which we simulate 25 showers. For a hexagonal array which is invariant for rotations of \SI{60}{\degree}, showers from \SI{0}{\degree} and \SI{30}{\degree} cover the two extreme cases of a shower falling into the array exactly parallel to a line of antennas and with the largest possible angle between two lines of antennas. Pulses are simulated on a grid with \SI{250}{m} spacing and a maximum axis distance of \SI{2235.6}{m} around the core. This amounts to $\lesssim 1350$ pulses per shower. The core location relative to a central antenna is randomly distributed. The pulses are simulated on a horizontal plane with an altitude of \SI{1400}{m} above sea level at its center\footnote{Unlike for the simulation set with the \SI{1.5}{km} detector layout, the detector plane with the dense \SI{250}{m} grid does not follow the Earth's curvature}. To study the reconstruction performance for different array spacings we define various sub-arrays. The following arrays are investigated: \SI{250}{m}, \SI{500}{m}, \SI{750}{m}, \SI{1000}{m}, \SI{1250}{m} and \SI{1500}{m}. For each spacing (except \SI{250}{m}) several unique sub-arrays can be defined, each of them corresponding to a different (relative) core position for a given simulation. Thus, for example, one single shower can be reconstructed on 36 unique sub-arrays with a spacing of \SI{1500}{m}. In Fig.\ \ref{fig:array} an example shower measured with the full \SI{250}{m} grid (top) and two different sub-arrays with a spacing of \SI{1500}{m} (middle, bottom) is shown. In Tab.\ \ref{tab:sim} the number of all unique sub-arrays for all 50 showers and the average number of antennas on these sub-arrays for each spacing are summarized. 

\begin{table}
  \caption{Average number of antennas simulated and maximum antenna-axis distance (measured perpendicular to the shower axis, i.e., in the shower plane) for the 1.5 km hexagonal grid as a function of the zenith angle in \SI{2.5}{\degree}-bins.}
  \centering\vspace{0.2cm}
  \begin{tabular}{c|cccccccc}
    \\[-1em]
$\langle \theta \rangle / ^\circ$ & 66.25 & 68.75 & 71.25 & 73.75 & 76.25 & 78.75 & 81.25 & 83.75 \\\hline
$\langle n_\mathrm{ant} \rangle \pm \sigma_\mathrm{ant} $ & 9 $\pm$ 1 & 10 $\pm$ 1 & 11 $\pm$ 1 & 16 $\pm$ 3 & 27 $\pm$ 6 & 47 $\pm$ 11 & 87 $\pm$ 21 & 173 $\pm$ 42 \\
$r_\mathrm{ant}^\mathrm{max}$ / m & 1500 & 1500 & 1508 & 1822 & 2230 & 2785 & 3563 & 4707
  \end{tabular}
  \label{tab:sim2}
\end{table}

\begin{table}
  \caption{The number of reconstructions $n_\mathrm{rec}$ performed on the dense simulations, i.e., the amount of all unique sub-arrays for all 50 showers, and average number of antennas on each sub-array $\langle n_\mathrm{ant} \rangle$ for the different array spacings.}
  \centering\vspace{0.2cm}
  \begin{tabular}{c|cccccc}
    \\[-1em]
    spacing / m & 250 & 500 & 750 & 1000 & 1250 & 1500 \\\hline
    $n_\mathrm{rec}$ & 50 & 200 & 450 & 800 & 1250 & 1800 \\
    $\langle n_\mathrm{ant} \rangle$ & 1342 & 336 & 149 & 84 & 54 & 37
  \end{tabular}
  \label{tab:sim}
\end{table}

\section{Interferometric reconstruction of the shower properties}
\label{sec:rit}
\begin{figure}[tbp]
    \centering
    \includegraphics[width=.35\linewidth]{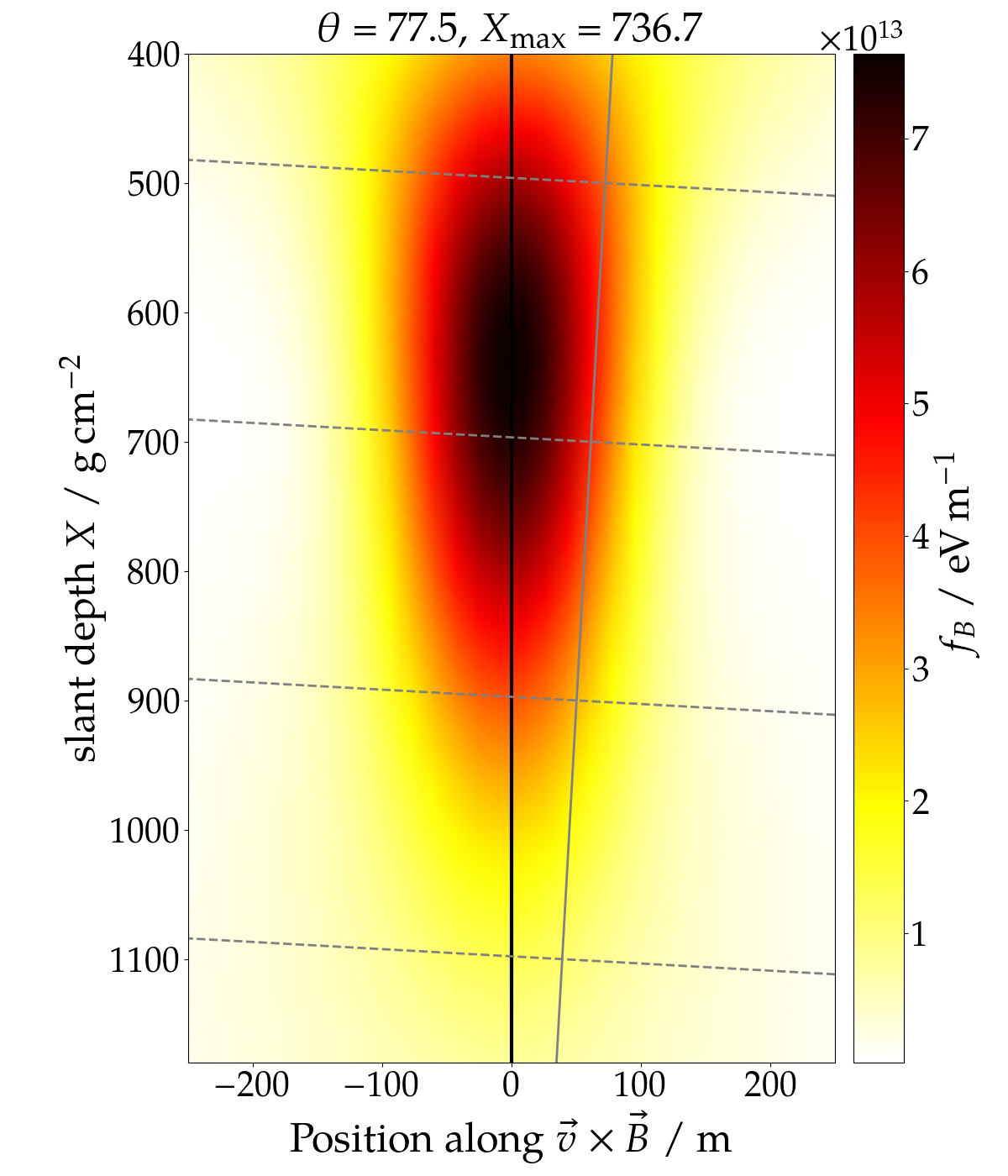}\hfill
    \includegraphics[width=.65\linewidth]{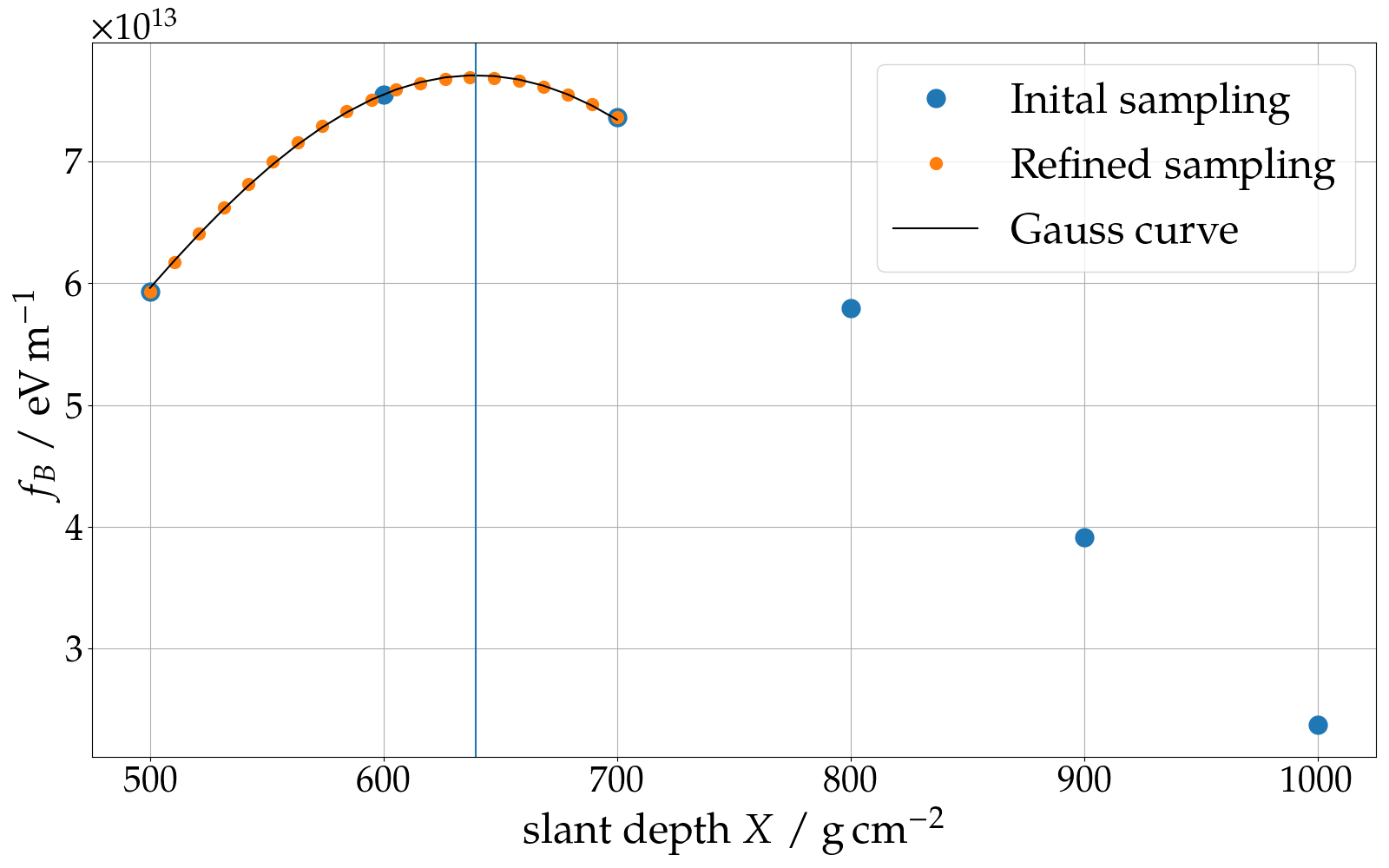}
    \caption{{\it Left}: Cross section of the coherent energy fluence profile of a \SI{1}{EeV}, \SI{77.5}{\degree} proton shower sampled with 1335 antennas on a 250 m grid. The cross section shows the longitudinal profile along the y-axis in g/cm$^2$ and the lateral profile in $\vec{v} \times \vec{B}$-direction along the x-axis. The coherent energy fluence is color-coded. The vertical black line indicates the shower axis. The grey lines illustrate 2-dimensional lateral cross-sections (dashed lines) along an initial guessed axis (solid grey line) used for the shower axis reconstruction (the shown line does not reflect the typical deviation of a guessed axis with a Gaussian resolution in $\phi$ and $\theta$ of \SI{0.5}{\degree} from the true shower axis). {\it Right}: Longitudinal profile of the coherent energy fluence $f_{B_{j}}$ for the same shower. First, to narrow down the position of the maximum, the longitudinal profile is sampled with a coarse \SI{100}{g\,cm^{-2}} sampling (large blue circles). Around the found maximum, in a \SI{200}{g\,cm^{-2}} window, the profile is then sampled more precisely with \SI{10}{g\,cm^{-2}} steps (small orange circles). In this window, a Gaussian curve is fitted to the profile (black curve). The vertical blue line shows the found maximum as determined from the fitted Gaussian parameters.}
    \label{fig:tomo}
\end{figure}
In this section, we describe the reconstruction of the shower axis and the depth of the shower maximum \Xmax with RIT. The algorithms, developed in \cite{schoorlemmer2020radio}, make use of 3-dimensional interferometric maps providing information about the longitudinal development of air showers. From these maps the cosmic-ray properties, in particular the arrival direction and depth of the shower maximum \Xmax, can be inferred. The algorithms described below are adapted from \cite{schoorlemmer2020radio}, however, their actual implementation is independent and has, in parts, changed. 

RIT exploits the coherence in the radio emission from air showers and one searches for an imaginary point source for which the coherent signal becomes maximal. The time-dependent coherent (beam-formed) signal $B_{j}(t)$ originating at an arbitrary location in the atmosphere $\vec{j}$ is calculated by the sum over all time-shifted antenna signals $S_{i}(t - \Delta_{i,j})$ at positions $\vec{i}$ 
\begin{equation}
    \label{eq:rit}
    B_{j}(t) = \sum_i^{n_\mathrm{ant}} S_{i}(t - \Delta_{i,j}).
\end{equation}
The time shift between an antenna location $\vec{i}$ and the source location $\vec{j}$ is
\begin{equation}
    \label{eq:rit2}
    \Delta_{i, j} = \frac{d_{i,j} \cdot \overline{n_{i,j}}}{c}    
\end{equation}
with the geometrical distance $d_{i,j}$ and effective (averaged) refractive index $\overline{n_{i,j}}$ between the positions $\vec{i}$ and $\vec{j}$, and the vacuum speed of light $c$. That means, $ \Delta_{i,j}$ corresponds to the light propagation time between positions $\vec{i}$ and $\vec{j}$.
To calculate the effective refractive indices between source locations $\vec{j}$ and antenna positions $\vec{i}$ a model for the refractivity $N(h) = n(h) - 1$ in the atmosphere is needed. For this analysis we adopt the Gladstone-Dale law together with a five-layer atmospheric density profile as used also in CoREAS in which the refractivity follows the density gradient with
\begin{equation}
    \label{eq:gladstone}
    N(h) = N(0) \cdot \rho(h) / \rho(0).
\end{equation}
In \cite{Mitra_2020} it is shown that this approximation is adequate for the frequency band of \SIrange{30}{80}{MHz}, for higher frequencies the Global Data Assimilation System (GDAS) can be used to refine the refractivity model by then also including the influence of humidity. The practical calculation of the effective refractivity between two positions, $\overline{N_{i,j}}$, which cannot be calculated analytically in a curved atmosphere, is explained in appendix \ref{sec:refracmodel}. The calculation of the light propagation time (using the effective refractivity) along straight lines corresponds to the algorithm adopted in CoREAS. In nature, the emission between sources and observers propagates on slightly bent trajectories due to refraction in the atmospheric refractive index gradient. In \cite{Schlueter_2020} we found that the calculation on straight lines reproduces the relative propagation times between two different sources in the atmosphere better than within \SI{0.1}{ns}, which is accurate enough to keep coherence properties in the frequency regime below a couple of hundred MHz.

To calculate $B_j(t)$, the electric field values of the time-shifted signals $S_i(t - \Delta_{i,j})$ are linearly interpolated to fit the finite time binning $\Delta t$ of $B_j(t)$\footnote{A linear interpolation is not strictly physically correct. Application of a phase gradient to the Fourier spectrum or adequate up-sampling of $S_i$ would be more physically motivated. However, linear interpolation is computationally more efficient and we validated that the reconstruction accuracy is independent of this procedure for $\Delta t = 1\,$ns for frequencies up to 200~MHz and $\Delta t = 0.33\,$ns for frequencies up to 350~MHz.}. For each trace $B_j(t)$ we determine a time-independent signal, namely the sum over the squared amplitudes in a \SI{100}{ns} signal window around the peak amplitude
\begin{equation}
    \label{eq:rit3}
    f_{B_{j}} = \epsilon_0 \, c \, \Delta t \sum_{t_\mathrm{peak}-50\mathrm{ns}}^{t_\mathrm{peak}+50\mathrm{ns}} B^2_j(t)
\end{equation}
where $\epsilon_0$ is the vacuum permittivity and $c$ the speed of light in vacuum. The peak amplitude and the peak time $t_\mathrm{peak}$ in $B_j(t)$ are determined from the maximum of the absolute Hilbert envelope of $B_j(t)$. The quantity $f_{B_{j}}$ can be understood as the coherent energy fluence received by the array of observers $i$ from a given location $\vec{j}$.

Eqs.\ \eqref{eq:rit}, \eqref{eq:rit2}, and \eqref{eq:rit3} allow us now to calculate the coherent energy fluence received from any position in the atmosphere. In Fig.\ \ref{fig:tomo} ({\it Left}) a cross section of the coherent energy fluence from an example shower sampled at 1335 antenna locations is shown. The longitudinal profile along the shower axis (vertical black line) is expressed in g$\,$cm$^{-2}$ (y-axis) while the lateral profile is shown perpendicular to the shower axis along the $\vec{v} \times \vec{B}$-direction (x-axis) in meters. It is apparent that the profile of the coherent energy fluence correlates with the particle cascade of the air shower, i.e., $f_B$ is strongest around the shower axis and exhibits a maximum. It has been shown that this maximum, defined as $X_\mathrm{RIT}$, correlates linearly with the shower maximum of the particle cascade \cite{schoorlemmer2020radio}. Thus RIT allows to reconstruct the shower properties, e.g., the depth of the shower maximum and shower axis. 

As in \cite{schoorlemmer2020radio}, only the signal in the $\vec{v} \times \vec{B}$ polarisation ($\vec{v}$: direction of the primary particle trajectory, i.e., shower axis, $\vec{B}$: direction of the Earth's magnetic field), which is obtained by rotating the Electric field vector simulated in the North-South, West-East, Vertical polarisations using the true arrival direction, is used for reconstruction ($\equiv S_i(t)$). It seems natural to separate the radio emission based on its emission mechanisms, i.e., separate between geomagnetic and charge-excess emission, as any phase-shift in the signals between both mechanisms would reduce the signals coherence. Such phase shifts correspond to a small degree of circular polarization observed both in simulations and data, see \cite{Huege:2016veh}. In inclined air showers the geomagnetic emission, which constitutes most of the signal in the $\vec{v} \times \vec{B}$ polarisation, is dominant while the signal in the $\vec{v} \times \vec{v} \times \vec{B}$ polarisation is completely comprised by the sub-dominant charge-excess emission. In fact, determining the longitudinal profiles $f_B(X)$ with signals in the $\vec{v} \times \vec{v} \times \vec{B}$ polarisation yields no well-defined maxima which can be correlated to the depth of the showers. 

\subsection{Reconstruction of the shower axis}
\begin{figure}[tbp]
    \centering
    \includegraphics[width=\linewidth]{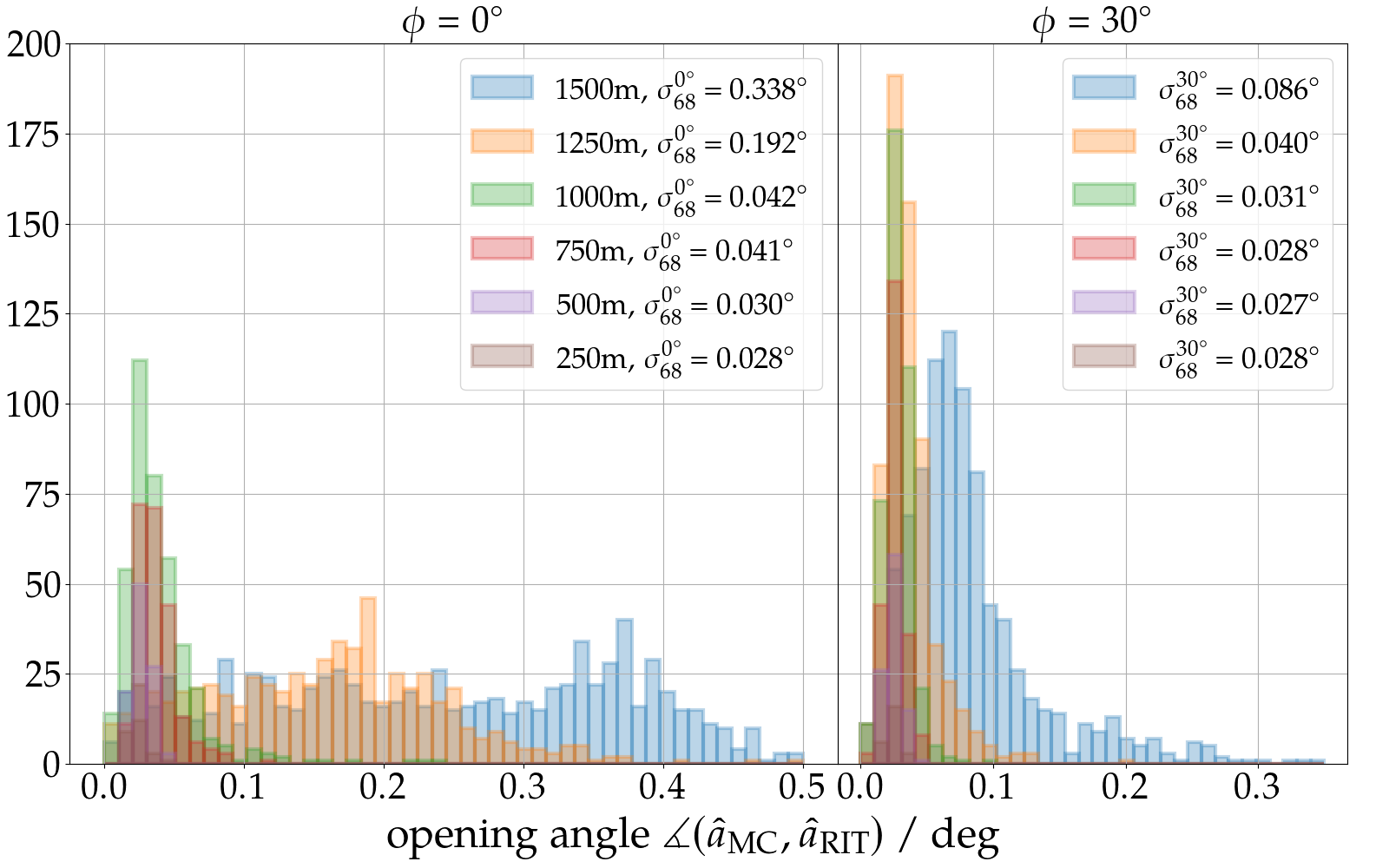}
    \caption{Histograms of the opening angle distribution between true and reconstructed arrival direction for the different antenna spacings (colors) for showers with $\phi = 0^\circ$ resp. $\phi = 30^\circ$. The legend shows the resolution of the arrival direction reconstruction in terms of the \SI{68}{\%}-quantile for all shower and all shower with $\phi = 0^\circ$ resp. $\phi = 30^\circ$.}
    \label{fig:axis}
\end{figure}

The shower axis, i.e., the extrapolated trajectory of the primary particle, is reconstructed with RIT by searching for an axis along which the longitudinal profile of the coherent energy fluence is maximal. For this propose, the lateral profile of the coherent radio emission, i.e., the cross-section of $f_{B_j}(X=\text{const})$, is sampled at several depths along the shower's development. For each cross-section the location of its maximum is determined and interpreted as its intersection with the shower axis. Given these intersections, a straight line is fitted minimizing the distance between line and intersections, weighted by the signal strength of each maximum.

Each cross-section is sampled in a plane perpendicular to an initial (guessed) axis which is determined given the true arrival direction, but smeared in zenith and azimuth angle with a Gaussian resolution of \SI{0.5}{\degree} each, and an intersection point at ground given by the intersection of the true Monte-Carlo (MC) shower axis smeared in a perpendicular plane with a Gaussian resolution of \SI{100}{m}. This accommodates for the imperfect knowledge of the shower axis from a traditional reconstruction
as starting point for a RIT reconstruction under practical circumstances. 
 
The following procedure is applied to find the maximum in each lateral cross-section at depths of 500, 600, 700, 800, 900, 1000 and \SI{1100}{g\,cm^{-2}}: In a first iteration, the maximum is searched on a quadratic grid which is characterized by its overall size and grid spacing. Here we chose a grid spacing of \SI{60}{m}. The grid covers an area of \SI{1}{km^2} and is set such that the location of the MC shower axis is within the search grid. This is due to performance reasons, not all MC shower axes would be contained in a \SI{1}{km^2}-grid around the initial guessed axis under the starting conditions mentioned above. The \SI{68}{\%} quantile of the distance between MC and guessed shower axis for an MC zenith angle of \SI{77.5}{m} at a depth of \SI{700}{g\,cm^{-2}} is \SI{675}{m}. However the area is sufficiently dimensioned to realistically model interferometric maps containing grating lobes, i.e., local maxima. For experimental measurements one has to ensure to make the search region sufficiently scaled, of course at the expense of computational effort. In a second iteration, the cross-section is sampled on a refined quadratic grid around the previously found maximum, i.e., zoomed-in around the previously found maximum. This process is repeated until the grid spacing becomes smaller than \SI{0.005}{\degree}.

In Fig.\ \ref{fig:axis} the opening angle distribution between true and reconstructed arrival direction for the dense simulations reconstructed on arrays with different antenna spacings with perfect time synchronisation is shown. The histogram is separated between showers arriving with azimuth angles of $\phi = 0^\circ$ ({\it Left}) and $\phi = 30^\circ$ ({Right}). The resolution in terms of the \SI{68}{\%}-quantile is shown in the respective legend. The overall accuracy, especially for antenna spacings $\leq 1000\,$m, is very good with less than $0.1^\circ$ for almost all configurations. For larger antenna spacings a bigger difference between showers from the two different incoming directions is evident. While the worsening of the resolution as function of the antenna spacing for showers from $\phi = 30^\circ$ is moderate and just becomes significant for the \SI{1500}{m} spacing, the degeneration for showers from $\phi = 0^\circ$ is much more dramatic. For those showers a footprint where all high-signal antennas are aligned on a straight line parallel to the shower axis projected on the ground is likely as the antenna grid gets too coarse to sample the Cherenkov cone along the whole plane (cf.\ middle panel of Fig.\ \ref{fig:array}). Inferring the correct arrival direction is more difficult for such geometries.

\subsection{Reconstruction of the shower maximum}
\begin{figure}[tbp]
    \centering
    \includegraphics[width=.85\linewidth]{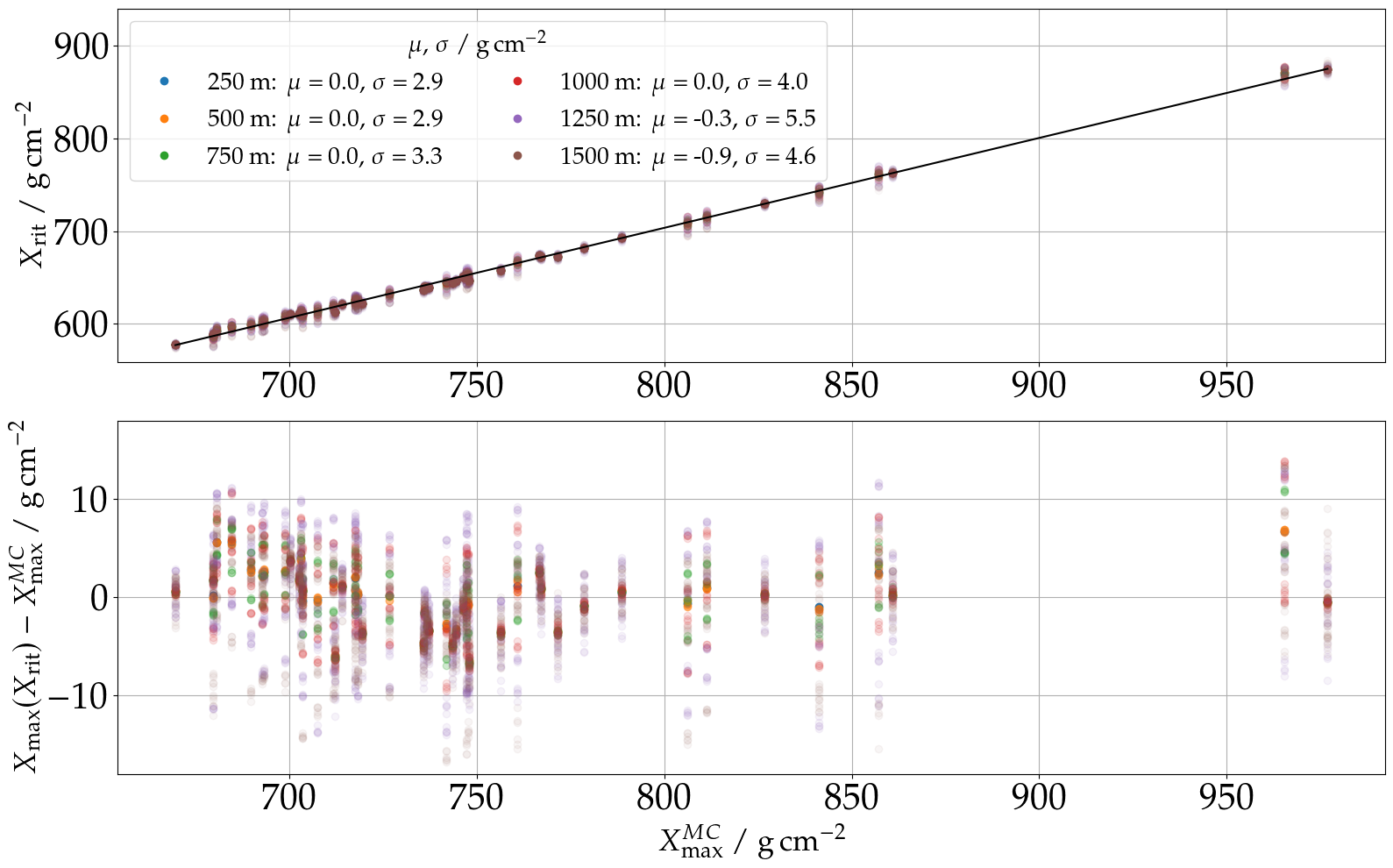}
    \caption{{\it Top}: Reconstructed $X_\mathrm{RIT}$ as a function of $X_\mathrm{max}^\mathrm{MC}$ for the dense simulations. The black line indicates the calibration curve according to Eq.\ \eqref{eq:xmax}. Reconstruction along the MC shower axis with perfect time synchronisation between the antennas. The different colors refer to reconstructions with (sub-)arrays of different spacings (transparency increases with number of reconstructions). The legend illustrates the bias and resolution for different array spacing. {\it Bottom}: Residuals between reconstructed and true depth of the shower maximum \Xmax.}
    \label{fig:aera1}
\end{figure}
To reconstruct the depth of the shower maximum, we determine the maximum $X_\mathrm{RIT}$ of the longitudinal profile of the coherent signal $f_{B_{j}}(X)$ along the (Monte-Carlo or reconstructed) shower axis. A profile of $f_{B_{j}}(X)$ along the MC shower axis as function of the slant depth $X$ is shown in Fig.\ \ref{fig:tomo} ({\it Right}). To find $X_\mathrm{RIT}$ we employ the following algorithm: We sample the longitudinal profile $f_{B_{j}}(X)$ along the shower axis in steps of \SI{100}{g\,cm^{-2}} between 500 and \SI{1000}{g\,cm^{-2}}. If the maximum is found at an edge, the sampling range is dynamically extended. Once the maximum is well-confined, a \SI{200}{g\,cm^{-2}} window around the found maximum is sampled with a refined step size of \SI{10}{g\,cm^{-2}}. $X_\mathrm{RIT}$ is then determined by the maximum of a Gaussian curve fitted to this \SI{200}{g\,cm^{-2}} window (cf.\ Fig.\ \ref{fig:tomo}, {\it Right}).

In Fig.\ \ref{fig:aera1} (top) the reconstruction of $X_\mathrm{RIT}$ for all dense simulations with a zenith angle of $\theta = 77.5^\circ$ and the different aforementioned array spacings (color coded) as a function of the true shower depth $X_\mathrm{max}^\mathrm{MC}$ is shown. The reconstruction is performed with a perfect time synchronisation between the different antennas, i.e., the signal arrival times are exactly known, and along the MC shower axis. A good, linear correlation is found between the reconstructed $X_\mathrm{RIT}$ and $X_\mathrm{max}^\mathrm{MC}$. Thus with a linear equation the shower maximum can be reconstructed as a function of $X_\mathrm{RIT}$:
\begin{equation}
\label{eq:xmax}
    X_\mathrm{max}(X_\mathrm{RIT}) = 1.03 \cdot X_\mathrm{RIT} + 76.15\,\text{g}\,\text{cm}^{-2}.
\end{equation}
The resulting residuals for the different spacings are shown on the bottom panel of the same figure (bias and resolutions of this residual in the legend of the top panel). It can be seen that regardless of the array spacing and thus the number of pulses used in the reconstruction (antenna  multiplicity) an accurate reconstruction is achieved.

\begin{figure}[tbp]
    \centering
    \includegraphics[width=\linewidth]{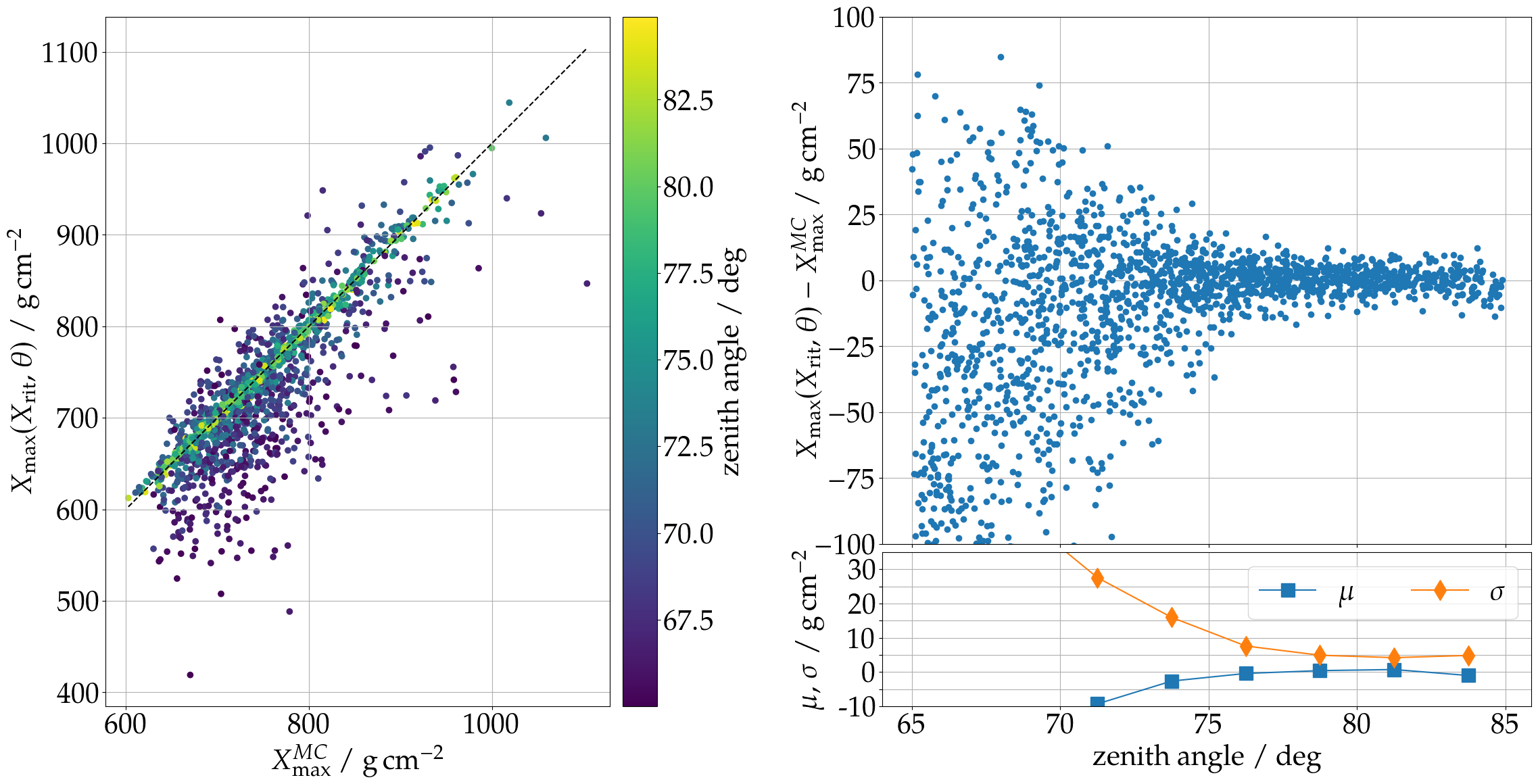}
    \caption{{\it Left}: Reconstructed \Xmax as a function of $X_\mathrm{max}^\mathrm{MC}$ for the simulations on the 1.5 km grid along the MC shower axis with perfect time synchronisation between the antennas. The reconstruction of $X_\mathrm{max}$ is a function of $X_\mathrm{RIT}$ and the zenith angle $\theta$ according to Eq.\ \eqref{eq:xmax2}. The black dashed line indicates the identity, the color code shows the zenith angle. {\it Right}: Residuals between reconstructed and true depth of the shower maximum \Xmax as function of the zenith angle. Bottom panel shows the profile, i.e., mean $\mu$ and standard deviation $\sigma$ of the above residual.}
    \label{fig:auger}
\end{figure}

In Fig.\ \ref{fig:auger} the reconstruction of \Xmax for the simulations on the \SI{1.5}{km} grid and showers with different zenith angles is shown. The application of a zenith-angle-independent calibration curve as in Eq. \eqref{eq:xmax} is insufficient. Introducing a simple linear zenith-angle dependency to the intercept parameter of Eq.\ \eqref{eq:xmax} is sufficient to accurately describe the relation between $X_\mathrm{max}$ and $X_\mathrm{RIT}$ for the here considered zenith angles range. A fit to showers with $\theta \geq 75^\circ$ yields the following calibration function:
\begin{equation}
    \label{eq:xmax2}
    X_\mathrm{max}(X_\mathrm{RIT}, \theta) =  1.04 \cdot X_\mathrm{RIT} + \left(68.31 - \frac{\theta - 77.5^\circ}{0.35^\circ} \right)\text{g}\,\text{cm}^{-2}.
\end{equation}
In Fig.\ \ref{fig:auger} ({\it Left}) the reconstructed \Xmax as function of the true $X_\mathrm{max}^\mathrm{MC}$ is shown. The comparison exhibits a significant scatter, only for showers with higher zenith angles (color coded) is a good correlation achieved.
The residual of the reconstructed \Xmax as function of the zenith angle and its profile (mean and standard deviation binned in \SI{2.5}{\degree} zenith-angle bins) is also shown ({\it Right}). It is apparent that the reconstruction accuracy strongly depends on the zenith angle. The dominant effect here is the insufficient antenna multiplicity for lower zenith angles (cf.\ Tab. \ref{tab:sim2}). The dependence of the reconstruction accuracy on the antenna multiplicity is investigated in more detail in section \ref{sec:ana}.

Comparing Eqs.\ \eqref{eq:xmax} and \eqref{eq:xmax2}, evaluated for $\theta = 77.5^\circ$, reveals no significant deviation between each other. Furthermore no significant bias between the \Xmax reconstructions with different antenna spacings is evident (cf.\ legend in Fig.\ \ref{fig:aera1}). Hence it seems that the calibration between \Xmax and $X_\mathrm{RIT}$ is independent of the antenna spacing and the different detector layouts covered in this work. The calibration found here is also in good agreement with \cite{schoorlemmer2020radio} which found an average depth $X_\mathrm{RIT} \sim 615\,$g$\,$cm$^{-2}$ for a true depth of the shower maximum of $X_\mathrm{max} \sim 700\,$g$\,$cm$^{-2}$ for showers with $\theta = 75^\circ$.

\subsection{Uncertainty due to atmospheric refractive index and density profiles}
\label{sec:atm}
In this section, we evaluate the impact of an inaccurately known atmospheric refractive index profile on the reconstruction. To that end, we reconstruct the showers with different  atmospheric refractivity profiles than used in the CoREAS simulations. We use the refractivity profiles ($\equiv$ refractivity at sea level \& atmospheric density profile) as adequate for the site of the Pierre Auger Observatory for the months of February and June for reconstruction, while the October profile was used for the CoREAS simulations. These two months represent the extrema in the yearly fluctuation of the refractivity at ground at the location of the Pierre Auger Observatory (for which the simulated October atmosphere resembles a good yearly average) \cite[Fig.\ 3.21]{PhDGlaser}. The yearly fluctuation is on the order of \SI{7}{\%} and thus larger than for the locations of other radio air-shower experiments such as LOFAR or Tunka-Rex with a yearly fluctuation of \SI{4}{\%} and \SI{3}{\%} \cite[p.~51]{PhDGlaser}, respectively. Thus using the refractivity profiles for February and June implies minimal knowledge of the true refractivity in the atmosphere.

\begin{figure}[tbp]
    \centering
    \includegraphics[width=.85\linewidth]{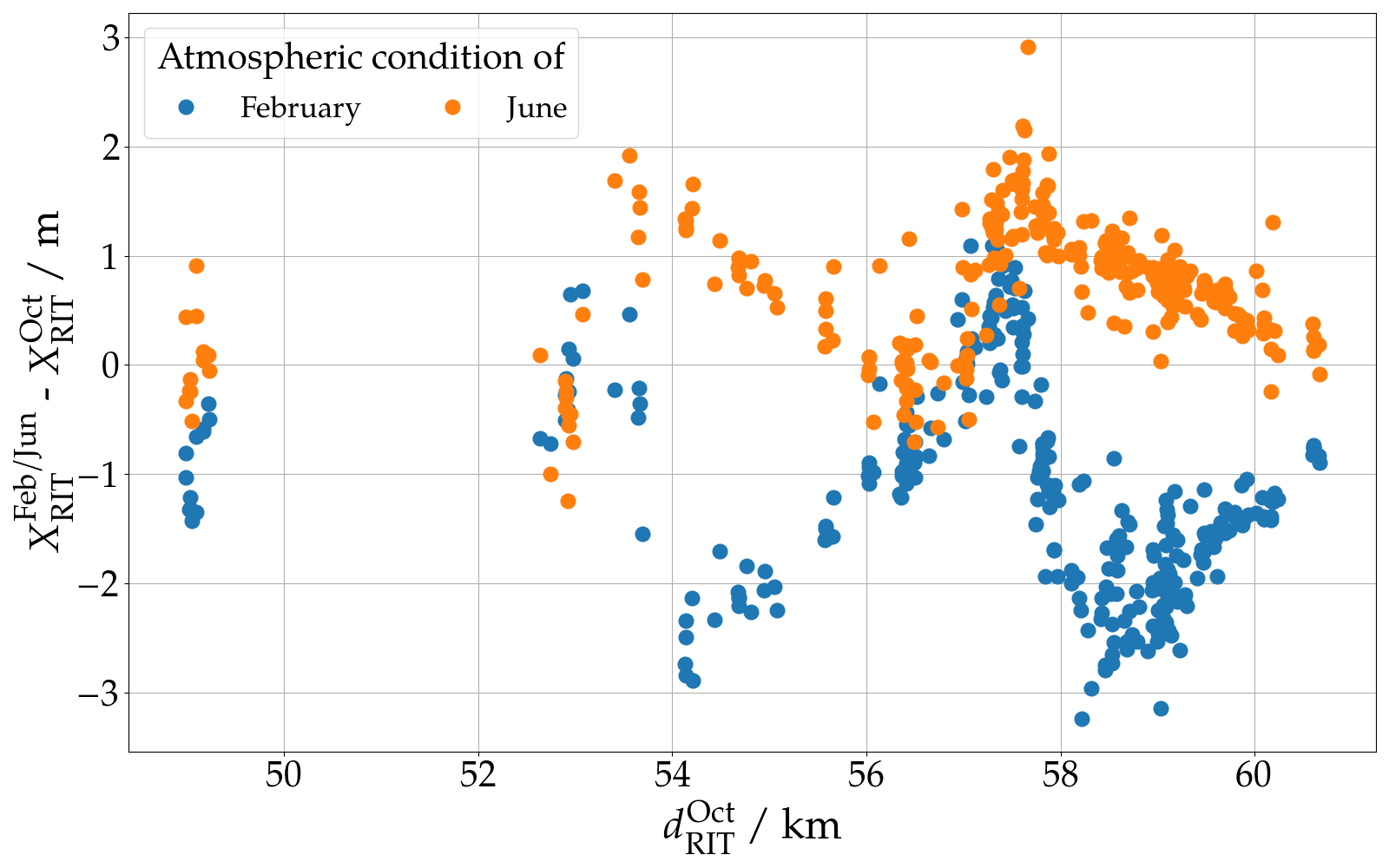}
    \caption{Deviation in $X_\mathrm{RIT}$ reconstructed with different refractive index profiles along the MC shower axis. A deviation of $\lesssim 3\,$g$\,$cm$^{-2}$ which cor response to $\lesssim 150\,$m is small compared to the absolute distance $\sim\,$\SI{55}{km}. Here, all 50 dense showers are reconstructed once for each array spacing.}
    \label{fig:refractivity_reco}
\end{figure}

Using a mismatching atmospheric density profile for reconstruction will yield a wrong atmospheric depth even if the point of origin of the maximally coherent emission is correctly determined. Thus, the atmospheric depth of a maximum, reconstructed with an inaccurate refractive index profile, is determined using the correct atmospheric density profile. The deviation in $X_\mathrm{RIT}$ between the reconstruction with different refractive index profiles shown in Fig.\ \ref{fig:refractivity_reco} is $\lesssim 3\,$g$\,$cm$^{-2}$. 

The uncertainty due to an inaccurate knowledge of the atmospheric density profile is identical to the corresponding uncertainty of \Xmax measurements with the fluorescence technique which is on the order of \SIrange{2}{4}{g\,cm^{-2}} \cite{PhysRevD.90.122005} for higher energies at the Pierre Auger Observatory.     

For the following investigation the (correct) simulated atmosphere density profile and refractive index at sea level is used for the reconstruction.

\section{Interferometric reconstruction of the depth of the shower maximum under realistic conditions}
\label{sec:ana}
Having evaluated the achievable performance under idealized conditions, largely confirming the results reported in \cite{schoorlemmer2020radio}, in the following sections we will evaluate the \Xmax reconstruction with RIT for a more practical scenario, i.e., with imperfect time synchronisation and along the reconstructed shower axis. 

\subsection{Reconstruction for a detector with a 1.5 km grid spacing}
\label{sec:auger}
Before we examine the interferometric reconstruction for simulations with varying detector layouts we evaluate the technique on simulations with the finite \SI{1.5}{km} antenna array and showers with varying zenith angles. In order to study the effect of imperfect time synchronisation between antennas we repeat the interferometric reconstruction several times after introducing random Gaussian time jitters. To gain quantitative insights, we evaluate the reconstruction quality in terms of the resolution (standard deviation) in \Xmax.

\begin{figure}[tbp]
    \centering
    \includegraphics[width=.85\linewidth]{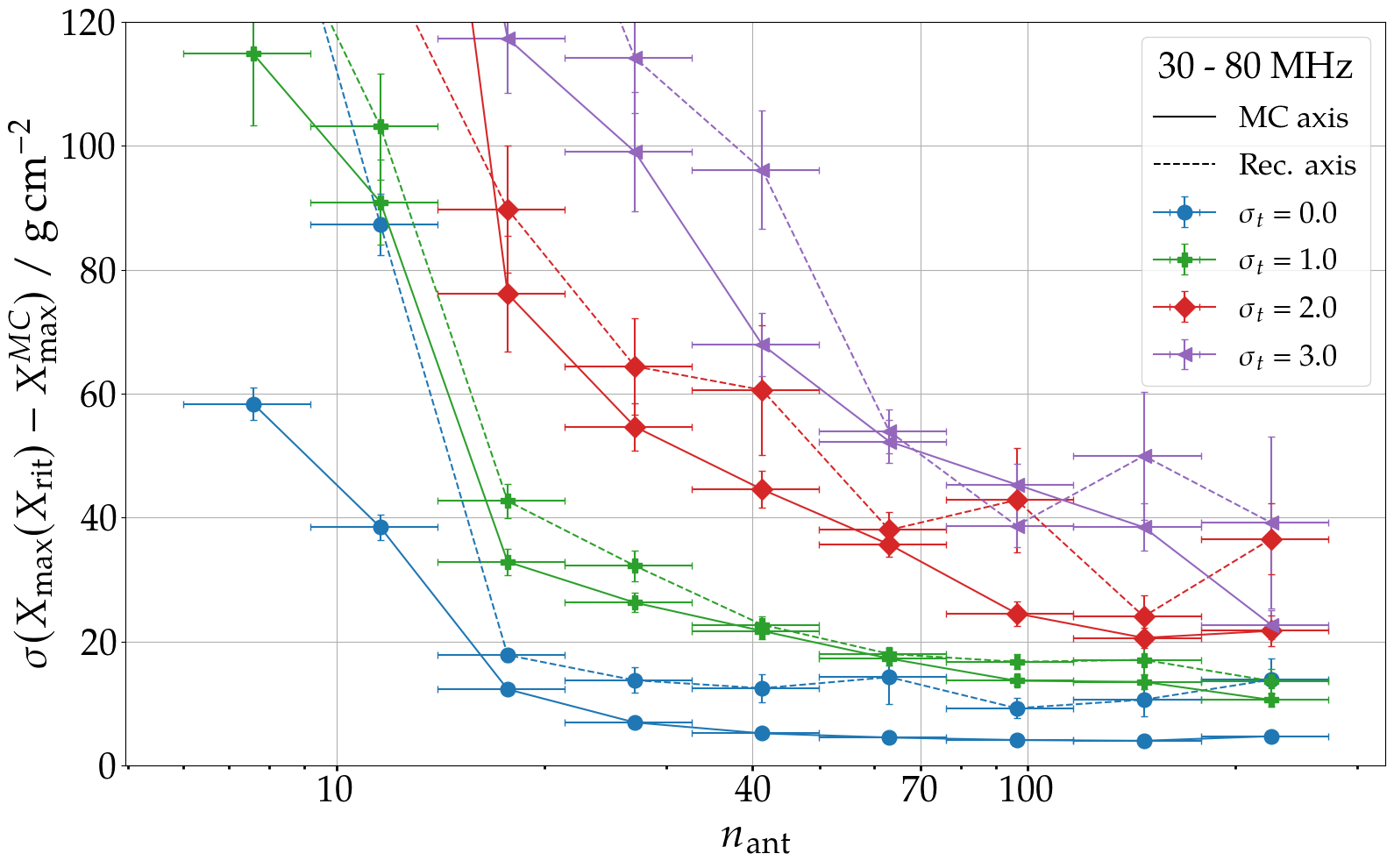}
    \caption{Resolution of the \Xmax reconstruction for the simulations on the 1.5 km grid along the true and reconstructed shower axes (solid and dashed lines, respectively), and for different Gaussian time jitters. The resolution is binned as a function of the antenna multiplicity, the horizontal error bars indicate the bin size, the vertical bars correspond to the statistical fluctuation of the resolution. The \Xmax reconstruction along the reconstructed axes can contain a few outliers where the axis reconstruction was inaccurate, causing the visible fluctuations in the \Xmax resolution.}
    \label{fig:auger2}
\end{figure}

In Fig.\ \ref{fig:auger2} the resolution of the \Xmax reconstruction via Eq.\ \eqref{eq:xmax2} binned as function of the antenna multiplicity is shown. The reconstructions along the MC and reconstructed shower axes are shown with solid and dashed lines, respectively. Different Gaussian time jitters are shown in different colors and markers. The horizontal error bars indicate the bin size, the vertical bars correspond to the statistical fluctuation of the resolution determined via a bootstrapping procedure. The reconstruction quality depends on both the antenna multiplicity and the accuracy of the time synchronisation. Even with a perfect time synchronisation, a minimum number of antennas $\gtrsim 12$ is required to keep the resolution below \SI{40}{g\,cm^{-2}}. An imperfect time synchronisation limits the achievable accuracy. Very accurate results ($\sigma_{X_\mathrm{max}} < 20\,$g$\,$cm$^{-2}$) are only achieved for a time jitter of \SI{1}{ns} or less and $\gtrsim 50$ antennas. It is visible that the effect of an increasing time jitter is more drastic for lower antenna multiplicities. This correlation is studied in more detail in the following section.

The reconstruction along the reconstructed axis exhibits a significant scatter, also for higher antenna multiplicities. This is due to a few but significant outliers which are result of mis-reconstructed axes (For example: For $\sigma_t = 2\,$ns, there is one outlier in the bin centered around $\langle n_\mathrm{ant} \rangle \sim 100$ and 3 in the last bin). 

In section \ref{sec:diss} the results obtained here are discussed and compared to other measurements of the depth of the shower maximum.

\subsection{Reconstruction for varying-density antenna arrays}
\label{sec:timejitter}
\begin{figure}[tbp]
    \centering
    \includegraphics[width=.85\linewidth]{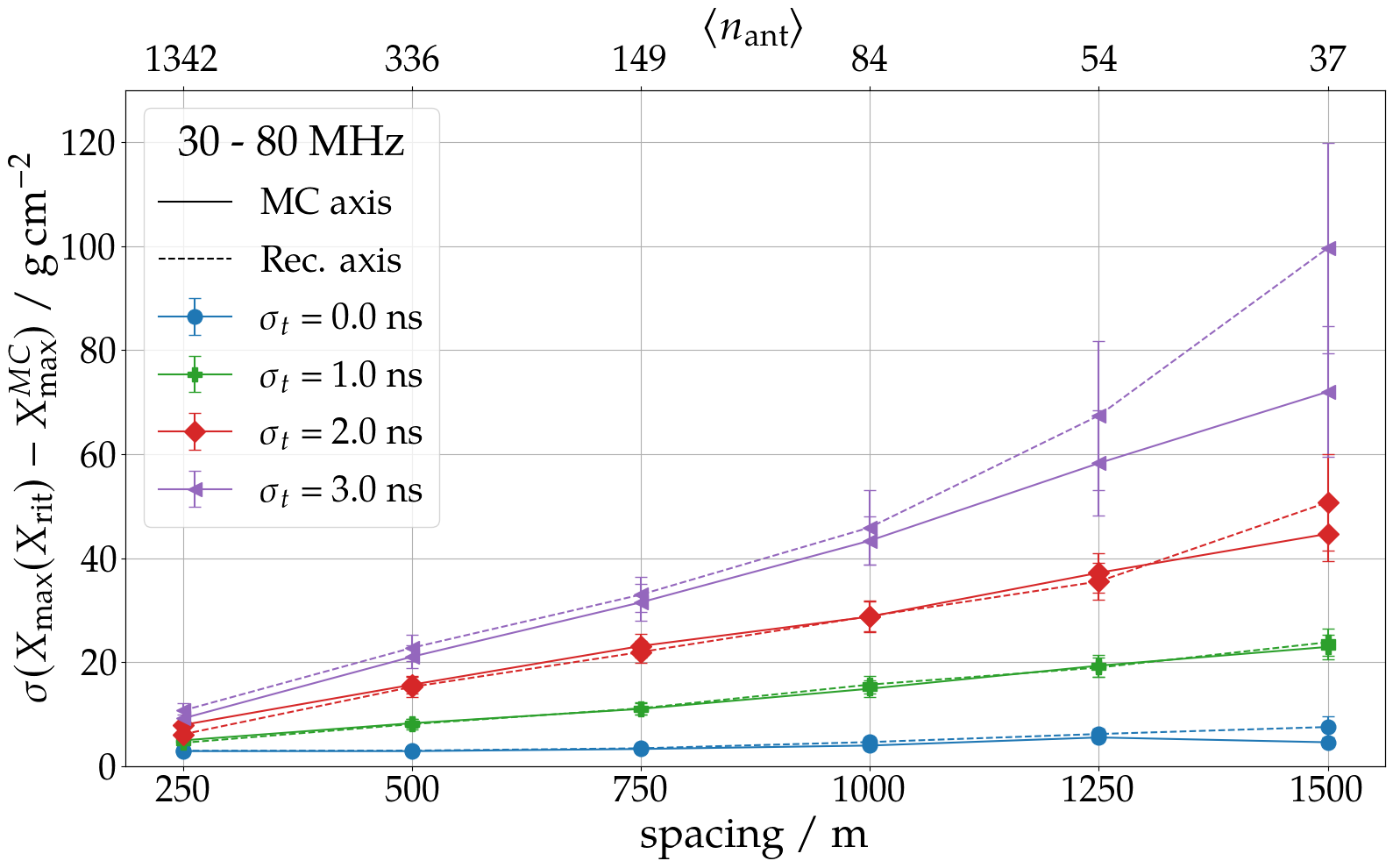}
    \caption{Reconstruction resolution in \Xmax of the 50 densely sampled showers with a zenith angle of \SI{77.5}{\degree}. Resolution is shown for different time jitter scenarios (different colors \& markers) and along the MC shower axis (solid line) or reconstructed axis (dashed line) as a function of the antenna spacing. The average number of antennas per spacing is given on the top x-axis as reference.}
    \label{fig:aera2}
\end{figure}

Now we study the effect of imperfect time synchronisation between antennas on the reconstruction of showers measured with different array spacings / antenna multiplicities. To that end, we repeat the reconstruction of the 50 simulated showers several times on various different sub-arrays after introducing random Gaussian time jitters mimicking an inaccurate time synchronisation between the antennas. Figure \ref{fig:aera2} shows the resolution in \Xmax as a function of the antenna spacing and for different time jitters. The average number of antennas $\langle n_\mathrm{ant} \rangle$ per event and spacing is shown on the top x-axis. The figure demonstrates that again the resolution worsens in the presence of a time jitter. This deterioration is amplified for showers reconstructed with a low antenna multiplicity. With a time jitter of \SI{3}{ns} the reconstruction on a very dense array with $> 1000$ antennas is still very accurate with a resolution of $\lesssim \,$\SI{10}{g\,cm^{-2}}. However, when the antenna multiplicity is $\lesssim \,$100, i.e., the showers measured with the \SI{1000}{m} grid or larger, the resolution deteriorates significantly to $\gtrsim \,$\SI{40}{g\,cm^{-2}}. Reconstructing \Xmax along an imperfectly reconstructed axis seems to have no significant implications for data taken with a Gaussian time jitter of \SIrange{0}{2}{ns} and only a little effect for the \SI{3}{ns} time jitter. In contrast to the simulations with the \SI{1.5}{km}-spaced array for which showers can be simulated at the edge of the finite \SI{3000}{km^2} array and thus their footprints are eventually not evenly sampled, the showers with the dense \SI{250}{m}-spaced array are always evenly sampled.  

Incorrect time synchronisation between antennas also affects the arrival direction reconstruction. In case of a \SI{3}{ns} Gaussian time jitter the resolution in the direction reconstruction worsens by a factor of $\sim$ 1.5 - 3 for all spacings. The difference in resolution between showers with $\phi = 0^\circ$ and \SI{30}{\degree} (cf.\ Fig.\ \ref{fig:axis}) decreases.

It is important to stress that the primary factor governing the \Xmax resolution, in case of imperfect time synchronisation between antennas, is the antenna multiplicity. The average number of antennas for each array spacing as listed in table \ref{tab:sim} refers to an instrumented area of \SI{72.5}{km^2} (cf.\ Fig.\ \ref{fig:array}). Smaller but denser arrays still need to accommodate a sufficiently high number of antennas and/or very good time synchronisation to allow accurate reconstructions. In addition, our tests have shown that a complete and symmetric sampling of the radio-emission footprint, i.e., inside, on top, and outside the Cherenkov cone, is needed to ensure accurate reconstruction. 

In section \ref{sec:diss} we use an analytic description of the radio-emission induced area as function of the zenith angle to generalize the results acquire here with showers with $\theta = 77.5^\circ$ and an instrumented area of \SI{72.5}{km^2} to lower zenith angles and smaller and denser arrays.

\section{Interferometric reconstruction for higher frequency bands}
\label{sec:freqs}
\begin{figure}[tbp]
    \centering
    \includegraphics[width=.75\linewidth]{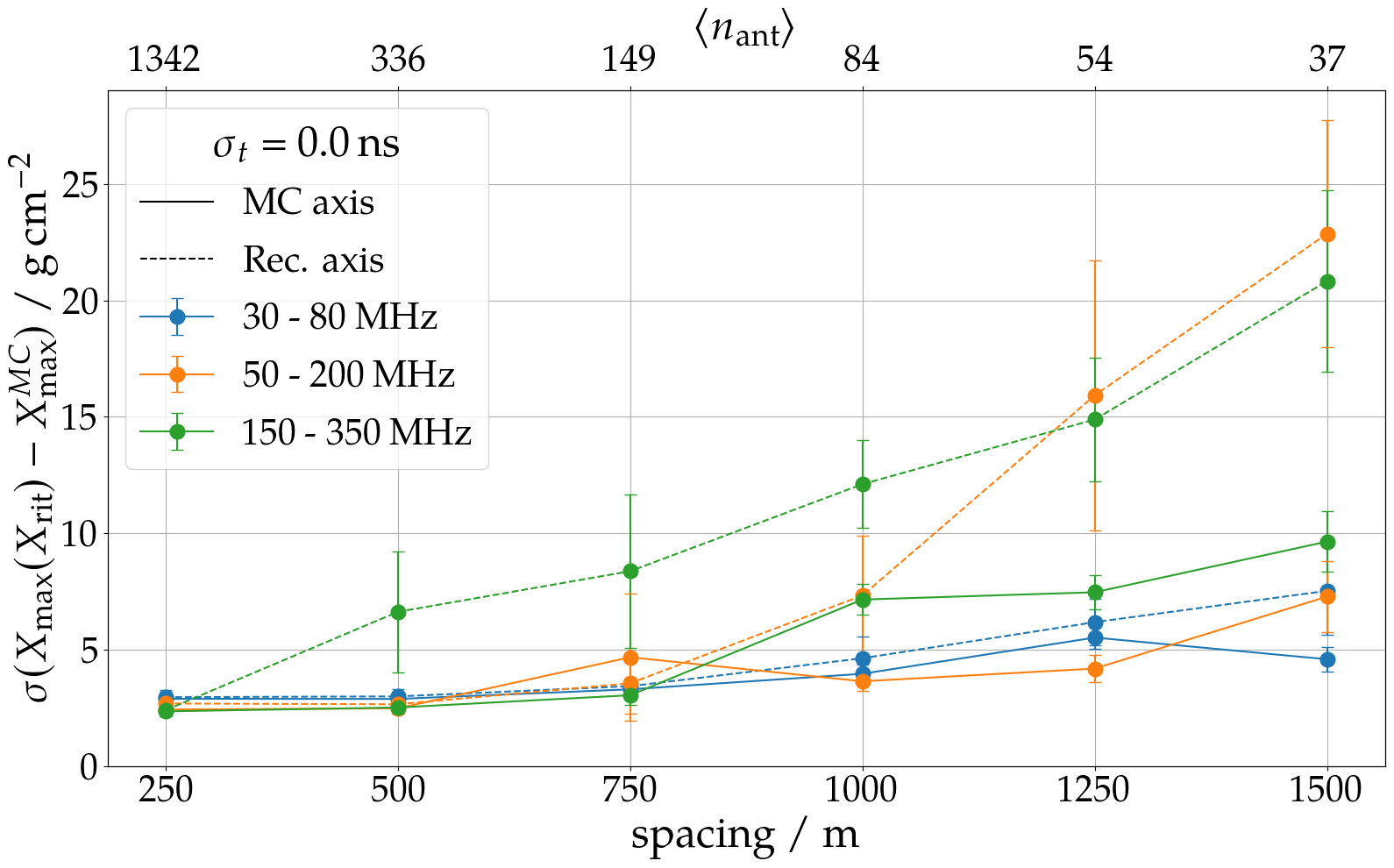}
    \caption{Reconstruction resolution in \Xmax for the \SIrange{30}{80}{MHz},  \SIrange{50}{200}{MHz}, and \SIrange{150}{350}{MHz} frequency bands (different colors) along the MC or reconstructed shower axis (solid or dashed line) for perfect time synchronisation as a function of the antenna spacing.}
    \label{fig:freqs1}
\end{figure}

Many next-generation radio-detection experiments aim to observe extensive air showers with broader frequency bands and at higher frequencies. Here, we test the interferometric reconstruction of \Xmax for two additional frequency bands: \SIrange{50}{200}{MHz} and \SIrange{150}{350}{MHz}. Applying the interferometric algorithm to data recorded at higher frequencies requires more stringent coherence criteria and thus a more accurate time synchronisation between antennas. The considered time jitter scenarios do not reflect equivalent phase-accuracy across the frequency bands but were chosen from a practical point of view, i.e., what time synchronisation accuracy an experiment has to achieve to employ RIT for higher frequencies. Furthermore, we found that the 3-dimensional profile of the coherent signal $f_{B_j}$ around the shower axis is increasingly narrow for higher frequencies. Hence the resolution with which the lateral cross-sections are sampled to infer the shower axis needs to be refined. For the axis reconstruction of showers recorded in the frequency band from \SIrange{150}{350}{MHz} an overall size of \SI{0.16}{km^2} and a search grid spacing of \SI{20}{m} was used.

\begin{table}
  \caption{Parameters of Eq. \eqref{eq:xmax}, i.e., $X_\mathrm{max} = a \cdot X_\mathrm{RIT} + b$ for the different frequency bands.}
  \centering\vspace{0.2cm}
  \begin{tabular}{c|cc}
    \\[-1em]
 & a & b \\\hline
 30 to 80$\,$MHz & 1.029 & 76.15$\,$g$\,$cm$^{-2}$ \\
 50 to 200$\,$MHz & 1.027 & 76.97$\,$g$\,$cm$^{-2}$ \\
 150 to 350$\,$MHz & 1.024 & 92.91$\,$g$\,$cm$^{-2}$
  \end{tabular}
  \label{tab:freq}
\end{table}

In Fig.\ \ref{fig:freqs1} the \Xmax resolution with perfect time synchronisation and along the MC and reconstructed shower axes for the different frequency bands (color coded) are compared. For the higher frequencies the reconstruction accuracy (along the MC shower axis) slightly decreases for sparser antenna arrays (\SIrange{150}{350}{MHz}). This is even more prominent for the reconstruction along the reconstructed axis (\SIrange{50}{200}{MHz} and \SIrange{150}{350}{MHz}).
One reason for this could be an insufficient sampling of the Cherenkov cone with these sparse arrays. As the Cherenkov cone itself is more dominating but also more narrow for showers measured with higher frequencies, the reconstruction accuracy depends more strongly on the antenna spacing. Furthermore, an offset in $X_\mathrm{RIT}$ between the different frequency bands is found, e.g., the $X_\mathrm{RIT}$ reconstructed for \SIrange{150}{350}{MHz} is $\sim \,$\SI{16.5}{g\,cm^{-2}} smaller as for \SIrange{30}{80}{MHz}. This offset has been taken into account when determining \Xmax. For this purpose we repeated the parametrisation of Eq.\ \eqref{eq:xmax} for the different frequency band, see Table \ref{tab:freq}. 

\begin{figure}[tbp]
    \centering
    \includegraphics[width=.49\linewidth]{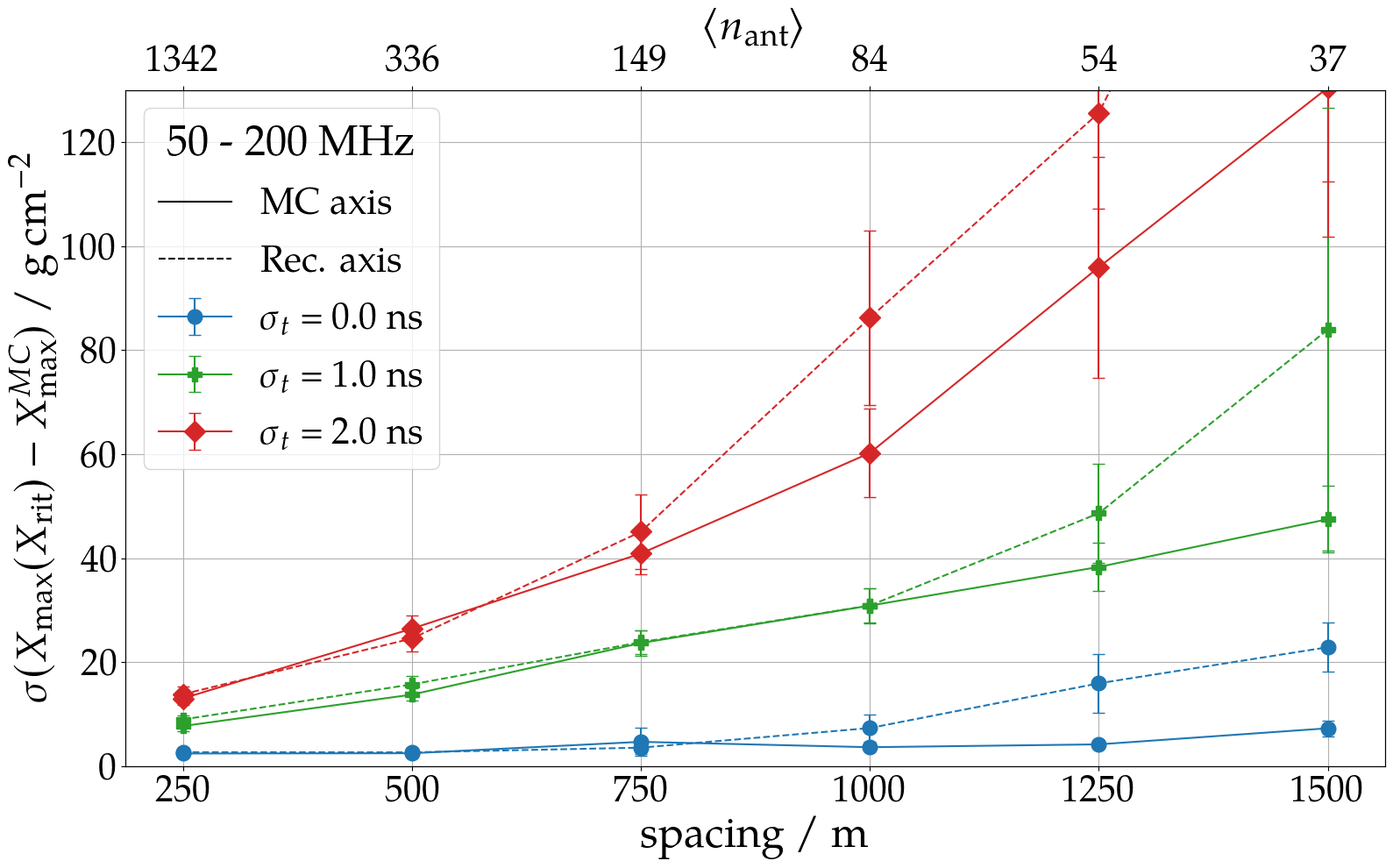}\hfill
    \includegraphics[width=.49\linewidth]{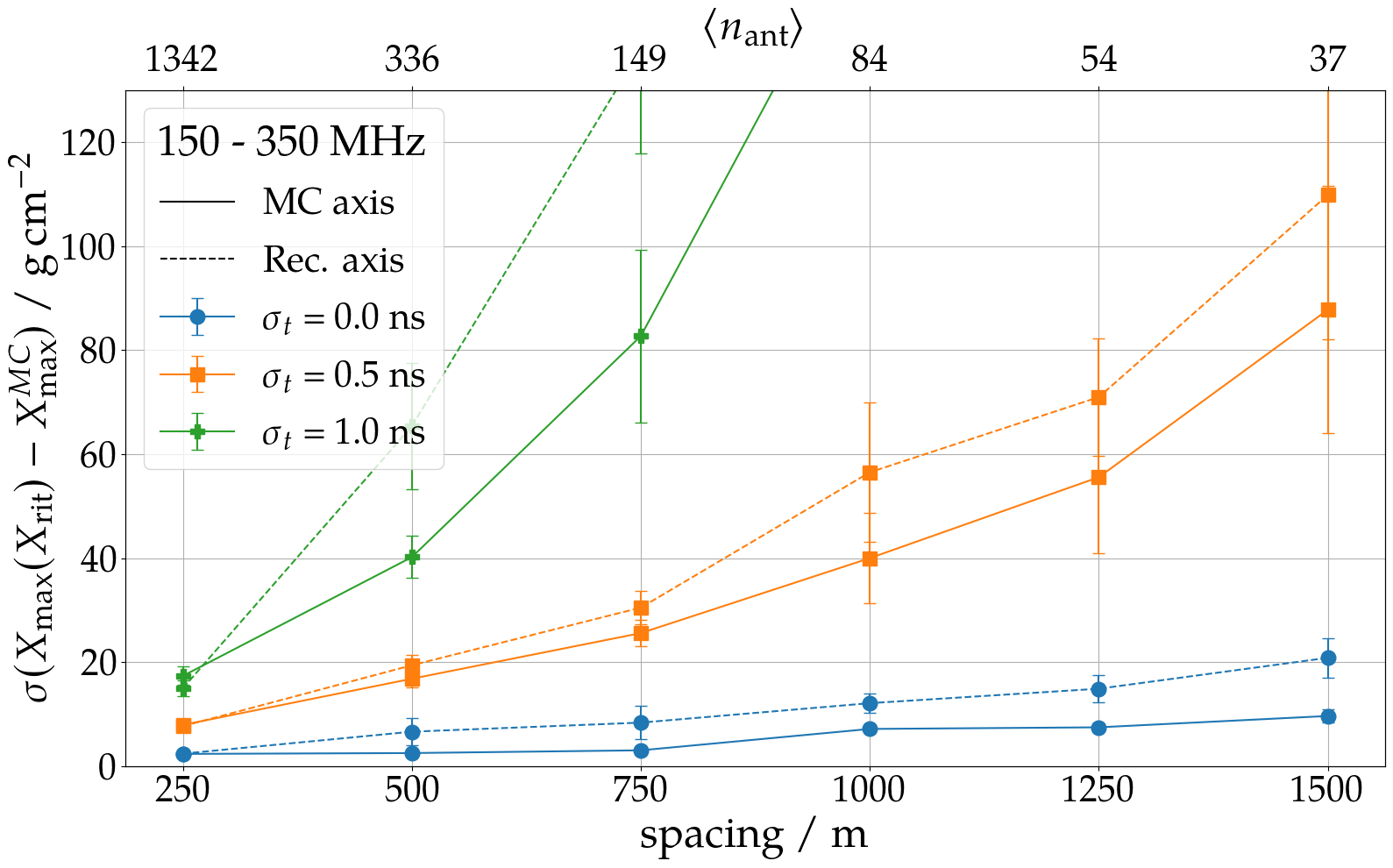}
    \caption{Reconstruction resolution in \Xmax for the \SIrange{50}{200}{MHz} ({\it Left}) and \SIrange{150}{350}{MHz} ({\it Right}) frequency bands and different time jitters (different colors and markers) along the MC or reconstructed shower axis (solid or dashed line) as a function of the antenna spacing.}
    \label{fig:freqs2}
\end{figure}

Figure \ref{fig:freqs2} shows the achieved \Xmax resolution for the two higher frequency bands ({\it Left}: \SIrange{50}{200}{MHz}, {\it Right}: \SIrange{150}{350}{MHz}) and different Gaussian time jitter scenarios. As expected, for the higher frequency bands already more modest time jitters, e.g., \SI{2}{ns} for \SIrange{50}{200}{MHz} and \SI{1}{ns} for \SIrange{150}{350}{MHz}, worsen the results significantly. The antenna multiplicity has to higher than $\sim \,70$ (100) and the time synchronisation better than $\sigma_t = 1\,$ns (0.5$\,$ns) for \SIrange{50}{200}{MHz} (\SIrange{150}{350}{MHz}) to achieve a resolution of $\sigma_{X_\mathrm{max}} < 40\,$g$\,$cm$^{-2}$.

As mentioned before, the axis reconstruction for \SIrange{150}{350}{MHz} requires a finer sampling of the lateral cross-sections. With the refined search-grid spacing of \SI{20}{m} a similar resolution of \Xmax, reconstructed along the reconstructed axis, compared to the lower frequency bands with a search-grid spacing of \SI{60}{m} is observed. However, the overall grid size used to reconstruct the shower maximum for data recorded with \SIrange{150}{350}{MHz} does not reflect the reconstruction under practical circumstances as the searched area is too small to reliably contain the true maximum for an axis (starting point) which is known with a accuracy of \SI{0.5}{\degree} in zenith and azimuth each.  

A, more sophisticated, gradient-descent based algorithm could reduce the computing time significantly and would allow the shower axis reconstruction under practical circumstances also for higher frequency bands. Such an algorithm has to be robust against grating lobes, i.e., local maxima in the interferometric maps.

In \cite{schoorlemmer2020radio}, in addition to an algorithm similar to the one described above, the shower axis is refined by maximizing the integrated longitudinal profile along the axis. Also in LOPES the arrival direction is inferred by a two-folded approach, first applying a raster-search algorithm and upon this a gradient-descent algorithm.

\section{Discussion}
\label{sec:diss}
Here, we discuss our results and also mention a few aspects which could not be studied in detail within the scope of this work. 

\begin{figure}[tbp]
    \centering
    \includegraphics[width=.85\linewidth]{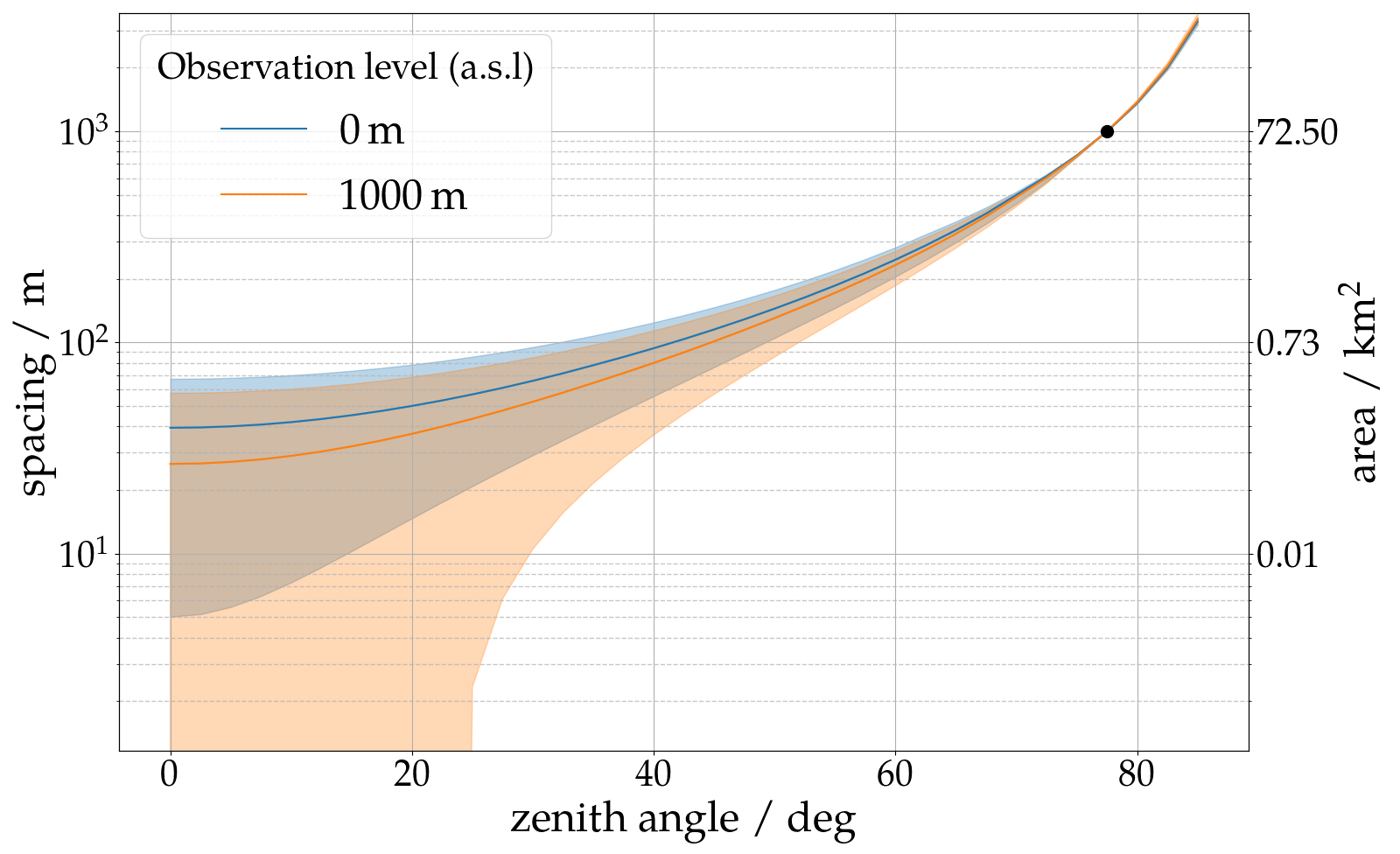}
    \caption{Antenna spacing required to achieve an antenna multiplicity of $\sim \,$84. This is the mean antenna multiplicity for showers with $\theta$ = \SI{77.5}{\degree} measured with a \SI{1000}{m} hexagonal grid (black dot) over an instrumented area of \SI{72.5}{km^2}. The second y-axis shows the area associated to a given antenna spacing and the aforementioned antenna multiplicity. For these showers an \Xmax resolution below \SI{20}{g\,cm^{-2}} when measured with a \SI{1}{ns} Gaussian time jitter in the \SIrange{30}{80}{MHz} band is achieved (cf.\ Fig.\ \ref{fig:aera2}). The different colors refer to different observation heights and the shaded areas correspond to \Xmax values ranging from \SIrange{550}{950}{g\,cm^{-2}} with a nominal value of \SI{750}{g\,cm^{-2}} (solid lines). The curves are calculated using Eqs.\ \eqref{eq:size} and \eqref{eq:radius}. For the calculation of the radius of the Cherenkov cone the US standard model of the atmospheric density profile and a refractivity at ground of $N_0 = 292 \cdot 10^{-6}$ are used. For showers with a zenith angles below \SI{25}{\degree} measured at an observation altitude of \SI{1000}{m} the shower maximum can lie underground, thus no lower limit can be calculated.}
    \label{fig:size}
\end{figure}

The study presented here for different detector layouts, especially those with an antenna spacing of < \SI{1.5}{km}, is limited to simulations with a zenith angle of $\theta = 77.5^\circ$. In \cite{schoorlemmer2020radio}, a modest improvement of the reconstruction accuracy with increasing zenith angle, i.e., with increasing distance between observer and source region, is found for simulations with a constant antenna multiplicity. The drastic improvement in resolution with increasing zenith angle found in this work (cf.\ Fig.\ \ref{fig:auger}) cannot be exclusively explained by this. In fact, the antenna multiplicity is identified as the crucial factor for an accurate reconstruction. Showers with a lower zenith angle illuminate smaller areas at ground and thus the antenna multiplicity decreases for showers measured with a constantly-spaced array. Assuming that the size $A$ of the radio-emission footprint at ground scales with the radius of the Cherenkov cone $r_\mathrm{che}$ (defined in the shower plane) yields the following relation:
\begin{equation}
\label{eq:size}
    A \sim \frac{\pi r_\mathrm{che}^2}{\cos \theta}.
\end{equation}
If an air shower is approximated as a point source moving with the speed of light, we can approximate the Cherenkov radius with the radius of a cone with its apex at the shower maximum and an opening angle equal the Cherenkov angle at the location defined by the refractive index $n(h)$, $\cos( \alpha_\mathrm{che}) = 1 / n(h)$:
\begin{equation}
\label{eq:radius}
    r_\mathrm{che} = \tan\left[\cos^{-1}\left(\frac{1}{n(h_\mathrm{max}(\theta, X_\mathrm{max}))}\right)\right] \cdot d_\mathrm{max}(\theta, X_\mathrm{max}, h_\mathrm{obs})
\end{equation}
with the height of the shower maximum above sea level $h_\mathrm{max}$, the distance along the shower axis between ground and shower maximum $d_\mathrm{max}$ and the altitude of the observation plane $h_\mathrm{obs}$. The antenna multiplicity is proportional to the footprint area $A$. To instrument a given area with a certain number of antennas the antenna spacing $\Delta_\mathrm{ant}$ has to satisfy the relation: $\sqrt{\Delta_\mathrm{ant}} \sim A$. Figure \ref{fig:size} shows the antenna spacing as a function of the zenith angle necessary to satisfy the antenna multiplicity $\langle n_\mathrm{ant} \rangle = 84$, i.e., the antenna multiplicity of the reference point at $\theta = 77.5^\circ$ (black marker) for showers measured with a \SI{1000}{m} hexagonal grid (cf.\ Tab.\ \ref{tab:sim}). Ignoring any additional zenith-angle related effects, this curve indicates the necessary antenna spacing to achieve a reconstruction as accurate as for the reference, i.e., $\sigma_{X_\mathrm{max}}(\sigma_t=1\,\text{ns}) = 16\,$g$\,$cm$^{-2}$ or $\sigma_{X_\mathrm{max}}(\sigma_t=2\,\text{ns}) = 29\,$g$\,$cm$^{-2}$ (cf.\ Figs.\ \ref{fig:auger2}, \ref{fig:aera2}). The different colors refer to different observation heights and the shaded areas correspond to \Xmax values ranging from \SIrange{550}{950}{g\,cm^{-2}} with a nominal value of \SI{750}{g\,cm^{-2}} (solid lines). These values resemble the mean and range of \Xmax values for a cosmic ray composition of half proton and half iron primaries with energies around \SI{10}{EeV}. It is apparent that an accurate reconstruction for more vertical showers with $\theta < 40^\circ$ is only achievable with antenna spacings below \SI{100}{m}. For showers with zenith angles below \SI{25}{\degree} the antenna spacing cannot be larger than $\sim$ tens of meters. Those showers, when measured at an altitude of \SI{1000}{m} a.s.l., can reach ground before developing the full maximum (this happens with a depth of the maximum of $> \,$\SI{850}{g\,cm^{-2}}) and thus no lower limit can be calculated. This emphasizes that for vertical air showers the observation altitude matters. This is further underlined by the finding in \cite{schoorlemmer2020radio} that the accuracy in \Xmax deteriorates when the distance between observer and source becomes smaller. The area associated to a given antenna spacing and $\langle n_\mathrm{ant} \rangle = 84$ is shown on the second (right) y-axis.

Furthermore, in this study the issue of triggering the readout of the radio signals, be it based on radio signals or measured particles, has not been considered. The interferometric reconstruction profits also from low signals and thus a readout of all antennas for a given event is optimal (as simulated in this study), even if no measurable radio pulse or particles are present. However most cosmic-ray experiments employ a trigger based on the signal strength per detector station to reduce the amount of data recorded. This might lead to a reduction of recorded radio pulses and thus limit the accuracy of interferometric measurements. Experiments with an accompanying particle detector can profit from a lower trigger threshold and thus recorded more radio pulses for vertical showers. For air showers with zenith angles beyond \SIrange{75}{80}{\degree}, however, the size of the radio-emission footprint eventually exceeds the size of the particle footprint \cite{Aab:2018ytv}, and a trigger relying on information of particle detectors alone will limit the amount of radio pulses recorded. For instance, the AugerPrime Radio Detector will only record radio pulses of antennas for which the water-Cherenkov detector (WCD) beneath has triggered. Thus the number of pulses recorded is governed by the particle footprint, i.e., the footprint for which the water-Cherenkov detectors will trigger, and as such is a function of the primary energy and the zenith angle. To trigger more than 12 WCD, a minimum zenith angle of $\gtrsim \,$\SI{75}{\degree} or energy of $\gtrsim \,$\SI{10}{EeV} has to be reached \cite{collaboration_2014}. More than 50 stations are almost never triggered. Given these limitations in the triggering for the AugerPrime Radio Detector, the application of a RIT reconstruction unfortunately does not seem very promising even if the time synchronisation can be improved to $\sigma_t \sim\,$\SI{1}{ns} (cf.\ Fig.\ \ref{fig:auger2}).

We point out that a description of the directional sensitivity of an appropriate radio antenna was not taken into account in this study. Moreover, we assume that the full 3-dimensional electric field vector, in particular the electric field in the $\vec{v} \times \vec{B}$ polarisation, is accessible from the experimental measurements. 

Besides detector effects, ambient noise is a crucial aspect for the detection of air showers with radio antennas. However, the ambient noise conditions can change dramatically between different locations around the Earth, their significance depends on the observed range of cosmic-ray energies and the frequency-band of choice, and their modeling taking into account different contributions, e.g., narrow- and broadband radio-frequency-interference, is not straightforward. Furthermore it is anticipated that the effect of ambient noise is attenuated for interferometric measurements scaling with the square root of the number of antennas \cite{schoorlemmer2020radio}. Dedicated studies for specific experiments are needed to determine the impact of noise at their specific location. Such a study can also determine if the interferometric detection threshold can be lowered when measuring in a higher frequency band \cite{V.:2017kbm}.

Judging the required \Xmax resolution to be achieved with a large sparse antenna arrays is a complex question as it depends on several factors such as the scientific objective, e.g., measuring the average mass composition or aiming for a light-heavy particle discrimination, the available statistics, and the astrophysical scenario, i.e., the actual mass composition of cosmic rays. To simplify, we compare the achievable \Xmax resolution with RIT to different experimental results. Recently, the Pierre Auger Collaboration has demonstrated that an accurate \Xmax reconstruction with the \SI{1.5}{km}-grid of water-Cherenkov (particle) detectors is possible using deep-learning techniques \cite{thepierreaugercollaboration2021deeplearning}. The resolution with this method for vertical showers with energies of around \SI{3}{EeV} is \SI{40}{g\,cm^{-2}} and improves to \SI{25}{g\,cm^{-2}} for energies above \SI{20}{EeV}. The resolution achieved by the Auger Fluorescence Detector is \SI{25}{g\,cm^{-2}} (\SI{15}{g\,cm^{-2}}) for energies above \SI{1}{EeV} (\SI{10}{EeV}) \cite{PhysRevD.90.122005}. LOFAR, a radio air shower experiment, measures vertical showers with hundreds of antennas in the energy range from 10$^{17}$ to 10$^{18}$~eV with a typical accuracy of \SI{17}{g\,cm^{-2}} \cite{PhysRevD.90.082003}. Tunka-Rex, another radio air shower experiment, measures \Xmax with a low, typical multiplicity of 7 antennas with an accuracy of \SI{25}{g\,cm^{-2}} \cite{PhysRevD.97.122004}. However, both results achieved with these radio experiments rely on the extensive use of very time-consuming and computing-intensive Monte-Carlo simulations and are not applicable for larger antenna arrays (with higher event statistics) or more inclined air showers.

The results obtained in this work show that the application of RIT with large, sparse antenna arrays relying on wireless communication is very challenging. Even with specialized hardware which improves the time synchronisation to $\sim \,$\SI{1}{ns} (or better for higher frequencies), an antenna multiplicity of $\gtrsim 50$ has to be achieved to obtain competitive results. These requirements will likely not be met by existing or currently planned experiments such as the Pierre Auger Observatory or GRAND.

More suitable for the application of RIT seem smaller ultra-dense antenna arrays with cabled communication such as the Square Kilometer Array SKA-Low \cite{Huege:2015jga}, which in fact is designed as an interferometer and thus will meet the required timing accuracy for interferometric analyses.

\section{Conclusion}
\label{sec:conc}
This study explores the potential for interferometric measurements of the depth of maximum of extensive air showers \Xmax with large arrays of radio antennas under realistic conditions. It has been shown that in addition to a very good time synchronisation of \SI{1}{ns} for the frequency band of \SIrange{30}{80}{MHz}, also a sufficiently large number of antennas per shower ($\gtrsim 100$) is needed for an accurate determination of \Xmax. Given the size of the radio-emission footprint as a function of the zenith angle, this constrains the maximal suitable array spacing to $< 100\,$m for vertical showers with zenith angles $\theta < 40^\circ$ and to a few hundred meters for showers with zenith angles $\theta \lesssim 75^\circ$. Only for higher zenith angles, arrays with an antenna spacing of \SI{1000}{m} or larger accommodate a sufficient antenna multiplicity. However, any kind of trigger based on the signal strength per detector station will reduce the amount of recorded radio pulses. Thus it seems very challenging to accommodate such requirements for (existing) air shower arrays with spacings $\gtrsim 1000\,$m which were designed and constructed without specific considerations for interferometric measurements, and in particular do not meet the requirements on the accuracy of time synchronisation.

The interferometric reconstruction of data recorded with higher frequencies showed no improvement in the achievable accuracy. Moreover we found that, in addition to more stringent requirements to the time synchronisation between antennas, no improvement in accuracy of the \Xmax reconstruction is achieved when the geometry and signals' arrival times are exactly known. Thus no advantage is found when applying the interferometric reconstruction to data recorded with higher frequencies.

Experiments which facilitate a large number of antennas combined with a very accurate time synchronisation such as the Square Kilometer Array SKA have great potential to exploit interferometric measurements of \Xmax. If combined with a muon detector, this approach could yield very valuable information to study the physics of extensive air showers and their hadronic interactions, as well as the mass composition of cosmic rays, with unprecedented detail.

\section*{Acknowledgements}
\emergencystretch=3em
We are grateful to F.G.\ Schröder for his valuable comments on our manuscript.
Felix Schlüter is supported by the Helmholtz International Research School for Astroparticle Physics and Enabling Technologies (HIRSAP) (grant number HIRS-0009). Simulations for this work were performed on the supercomputer BwUniCluster 2.0 and ForHLR II at KIT funded by the Ministry of Science, Research and the Arts Baden-Württemberg and the Federal Ministry of Education and Research. The authors acknowledge support by the state of Baden-Württemberg through bwHPC.

\section{Appendix}
\subsection{Calculation of the effective refractive index between two arbitrary locations in the atmosphere}
\label{sec:refracmodel}
From equation \eqref{eq:gladstone} follows that the effective refractivity between two positions $\vec{i}$ and $\vec{j}$ is calculated via the integral along the line of sight with length $l_{i,j}$:
\begin{equation}
    \label{eq:int}
    \overline{N_{i,j}} = \frac{N(0)}{\rho(0)} \frac{\int_i^{j} \rho(h(l)) \mathrm{d}l}{l_{i,j}}.
\end{equation}
For sources with a zenith angle $\theta \lesssim 60^\circ$ the atmosphere can be approximated to be flat and the integral over $\mathrm{d}l$ can be substituted with $\mathrm{d}l = \mathrm{d}h/\cos(\theta)$. This simplifies the equation to
\begin{equation}
    \overline{N_{i,j}} = \frac{N(0)}{\rho(0)} \frac{T_i - T_j}{\Delta h_{i,j}}
\end{equation}
with the analytically described mass-overburden $T_x = T(h_x)$. For more inclined geometries the curvature of the Earth has to be taken into account and the density along $l_{i,j}$ becomes a function of the zenith angle and distance from ground. In that case the integral in Eq.\ \eqref{eq:int} cannot be solved analytically anymore and thus has to be solved numerically. This is computationally very demanding and thus we use a pre-calculated table of the integrated refractivity\footnote{This is equivalent to the treatment included in CoREAS since version v7.7000.}. The table comprises the integrated refractivity as a function of the zenith angle a given line of sight makes with the Earth's surface (which is not identical to the zenith angle measured at higher altitudes along the line of sight) and the distance $d$ between the Earth's surface and a point along the line of sight. For any two points in the atmosphere for which the line through both points also intersects with the spherical Earth the integrated refractivity between those points can then be determined directly from the pre-tabulated values. The grid points of the table are spaced in $\tan\theta$ and equidistant in distance $d$. The integrated refractivity for any arbitrary point in the atmosphere is determined by a bi-linear interpolation within this table. A python implementation of this table and the interpolation has been made publicly available at \cite{radiotools}. 

\begin{figure}[tbp]
    \centering
    \includegraphics[width=.85\linewidth]{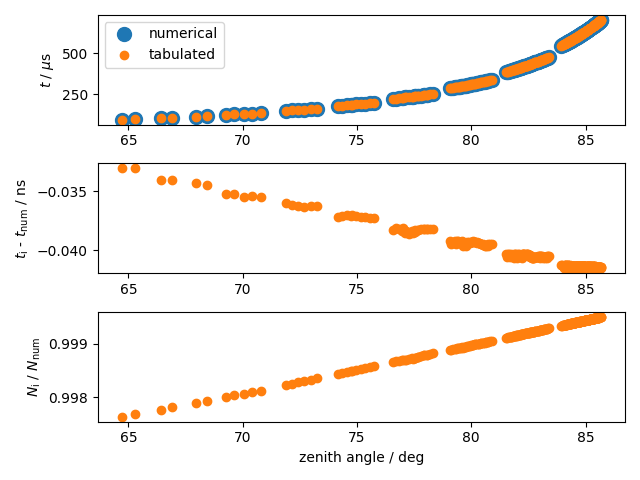}
    \caption{{\it Top}: Total light propagation time for (9) different source positions and numerous different observers. The x-axis denotes the zenith angle under which a source is seen by an observer. The propagation time is calculated along straight lines with the effective refractivity being calculated by very fine-grained piece-wise numerical integration or with the pre-calculated tables. {\it Middle}: Absolute difference of the light propagation time between numerical and tabulated calculation. {\it Bottom}: Relative agreement of the effective refractivity between numerical and tabulated calculation.}
    \label{fig:refractivity}
\end{figure}

In Fig.\ \ref{fig:refractivity} we compare the effective refractivity determined with this model (pre-calculated table) to the exact numerical solution. The total light propagation time can be accurately calculated using the tabulated effective refractivity (cf. Fig.\ \ref{fig:refractivity} top panel). A minor zenith-angle-dependent bias is visible in the absolute residual (cf.\ middle panel), however, the deviation is negligible since it is below any coherence criteria for signals in the MHz regime. In addition to the bias, small {\it wiggles}, which originate from the unpyhsical, linear interpolation between zenith angle bins, are visible but also negligible.

\clearpage

\bibliographystyle{unsrtnat}  
\bibliography{main}  

\end{document}